\documentclass[preprint,showpacs,prf,floatfix,longbibliography]{revtex4-1}

\usepackage[utf8]{inputenc}
\usepackage{amsmath}
\usepackage{amssymb}
\usepackage{lipsum}
\usepackage{bm}
\usepackage{graphicx}
\usepackage{graphics}
\usepackage[dvipdf]{epsfig}
\usepackage{subcaption}
\usepackage{float}
\usepackage{wrapfig}
\usepackage{mathtools}
\usepackage{url}
\usepackage[dvipsnames]{xcolor}

\begin{document}

\title{Model of the dynamics of an interface between a smectic phase and an isotropic phase of different density}

\author{Eduardo Vitral and Perry H. Leo}
\affiliation{Department of Aerospace Engineering and Mechanics,
    University of Minnesota, 110 Union St. SE, Minneapolis, MN 55455, USA}
\author{Jorge Vi\~nals}
\affiliation{School of Physics and Astronomy,
    University of Minnesota, 116 Church St. SE, Minneapolis, MN 55455, USA}

\begin{abstract}
Soft modulated phases have been shown to undergo complex morphological transitions, in which layer remodeling induced by mean and Gaussian curvatures plays a major role. This is the case in smectic films under thermal treatment, where focal conics can be reshaped into conical pyramids and concentric ring structures. We build on earlier research on a smectic-isotropic, two phase configuration in which diffusive evolution of the interface was driven by curvature, while mass transport was neglected. Here, we explicitly consider evaporation-condensation processes in a smectic phase with mass transport through a coexisting isotropic fluid phase, as well as the hydrodynamic stresses at the interface and the resulting flows. By employing the Coleman-Noll procedure, we derive a phase-field model that accounts for a varying density field coupled to smectic layering of the order parameter. The resulting equations govern the evolution of an interface between a modulated phase and an isotropic fluid phase with distinct densities, and they capture compressibility effects in the interfacial region and topological transitions. We first verify a numerical implementation of the governing equations by examining the dispersion relation for interfacial transverse modes. The inverse decay rate is shown to scale as $Q^{2}$ ($Q$ is the wavenumber of the perturbation) due to hydrodynamic effects, instead of the $Q^{4}$ expected for diffusive decay. Then, by integrating the equations forward in time, we investigate fluid flow on distorted layers and focal conics, and show how interfacial stresses and density contrast significantly determine the structure of the flow and the evolution of the configuration.
\end{abstract}

\date{\today}

\maketitle


\section{Introduction}

Soft matter systems are the subject of research in a number of areas due to the versatility of their ordered phases, and the easy control of morphology, defects and topology, leading to potential novel applications in both materials science and biology. Among prominent soft materials we mention modulated phases in block copolymers \cite{vu2018curvature} and smectic liquid crystals \cite{dierking2015smectic}. The latter are formed by anisometric molecules that present collective orientational order along a director axis, and are organized in periodically spaced layers, so that they exhibit broken rotational and translational symmetries. While in the longitudinal direction layers behave rigidly as a solid, the transverse direction (in the two dimensional manifolds defined by the smectic layers) exhibits fluidity, and hence both elasticity and hydrodynamics are important when modeling smectics. Both effects have a major role in the orientational control of smectics, a fact that has prompted many recent experimental studies of mesoscopic patterning of liquid crystal films by thermal and surface treatments \cite{kim2014creation,ryu2015creation,kim2016controlling}, and also by manipulating the geometry of the interface via inclusions \cite{beller2013focal}. This combination of elasticity and hydrodynamics becomes key not only to the engineering of surface properties, but also to controlling the structure of self-induced flows.

Continuum models of liquid crystals, including evolving two phase interfaces, are an important tool for understanding the rich interplay between hydrodynamics, topology, structure and curvatures during morphological transitions. Theoretical and numerical efforts in this direction have been made for nematic liquid crystals, initially for a fixed surface \cite{napoli2016hydrodynamic} and more recently for a moving nematic with a minimal continuous surface \cite{nitschke2019hydrodynamic}, coupling hydrodynamics with interfacial evolution. Particular interest lies in how the fluid flow connects to the director field and geometry of such systems, and the role played by hydrodynamics in the interaction of defects. This control is also key in studies of active matter transport, since active particles and microswimmers, such as bacteria, can be guided by the flow induced by an anisotropic medium \cite{daddi2018dynamics,ferreiro2018long}. However, a theory for the coupled evolution of a smectic liquid crystal with a two phase interface and the resulting hydrodynamic flows, including macroscopic singularities associated to topological defects, is still under active development. The periodic nature of smectic phases requires one to distinguish between the motion of the surfaces defining the modulation in the bulk phase, and the interface separating this phase from a neighboring isotropic fluid phase. When studying the evolution of such interfaces, the model must also be able to accommodate topological transitions and dynamically handle macroscopic singularities in the form of defects. Phase-field models, or Ginzburg-Landau-type equations as used in our work, present a versatile way to describe complex interfacial morphologies and their evolution, being able to represent a modulated to disordered transition by a smooth order parameter \cite{mukherjee2001simple,pevnyi2014modeling}. 

We previously introduced a diffuse interface model of a smectic-isotropic interface \cite{vitral2019role} with uniform density, and focused our attention on the resulting thermodynamic relations and kinetic laws in the macroscopic limit of a thin interface. While the model describes diffusive evaporation-condensation, a proper study of a smectic film interface requires consideration of mass flow and stresses at the interface as the smectic is in contact with an isotropic fluid phase of different density. For this, we need a complete model including conservation of momentum, mass, and a dynamic equation for the order parameter, with a density dependent free energy. We present here a detailed derivation of a phase-field model for a system where a smectic phase is in contact with an isotropic phase of different density. We further consider the quasi-incompressible limit in order to focus on mass transport at the interface, in an effort to model the experiments on smectic thin films of Refs. \cite{kim2016controlling,kim2018curvatures}.

Cahn and Hilliard \cite{re:cahn58} pioneered the use of phase-fields in the study of interfacial motion in a binary mixture by assuming a gradient free-energy functional of the concentration. The method was further developed to study the unstable motion of a two-phase interface by Allen and Cahn \cite{re:allen79}. The model was also extended to include hydrodynamic flows \cite{re:jasnow96,gurtin1996two} through a coupled Navier-Stokes and Cahn-Hilliard problem, also known as Model H in the critical dynamics literature. Lowengrub and Truskinovsky \cite{lowengrub1998quasi} derived a phase-field model for a binary mixture with phases of different density, and derived a thermodynamically consistent model that accounts for the effects of such a varying density field. While they considered both bulk phases of the binary to be incompressible, they show that compressibility effects take place at the interface, where the velocity becomes non-solenoidal. Since the density can be calculated by a constitutive equation from any point where the composition is known, and compressibility is restricted to the interface, their model is known as the quasi-incompressible Cahn-Hilliard model. Diffuse-interface models and numerical schemes for quasi-incompressible two-phase flows with distinct densities have been actively developed since then \cite{lee2002modeling,abels2012thermodynamically,guo2017mass,gong2018fully}, with particular interest in large density ratios \cite{ding2007diffuse,shokrpour2018diffuse}, for which the stability of the derived numerical schemes becomes a problem due to nonlinear terms coupled to the density. These latter developments may be extended beyond uniform bulk phase binaries. For the case of a modulated-isotropic interface, however, the oscillatory nature of the order parameter introduces challenges that we explore in the present work. For example, we have an energy density with a dependence on higher order derivatives, and a less intuitive choice of constitutive equation for the density. In this case, we need to extract from a non uniform oscillatory order parameter a density field that is homogeneous in both phases, and carefully define pressure fields.

Here we use a Coleman-Noll procedure \cite{tadmor2012continuum,lowengrub1998quasi,shokrpour2018diffuse} to derive a set of governing equations that couple the phase-field equation for the order parameter representation of a smectic-isotropic system to a momentum transport equation, and account for a varying density between the two phases. We specialize our discussion to the analog of quasi-incompressible smectic-isotropic fluid, motivated by recent experiments in smectic A thin films \cite{kim2016controlling,kim2018curvatures}. When such films are deposited on treated substrates, antagonistic boundary conditions cause smectic layers to align perpendicularly to the substrate but parallel to the interface with the fluid. In this way they induce the smectic layers to bend into focal conics, which are topological defects that organize throughout the film into periodic arrays. It has also been observed that when domains of focal conics are formed through morphological transformations in the nematic-smectic transition, they retain the geometric memory of how boojum defects in nematics were organized \cite{gim2017morphogenesis,suh2019topological}. By thermal annealing, these focal conics are reshaped into various other structures due to the curvature driven evaporation-condensation of the smectic layers, leading to a variety of morphologies, including conical pyramids and concentric rings. These structures present dual scale features, since the scale of the original defects is usually in micrometers, while the details of the formed layers are nano-sized. For example, this dual roughness gives them superhydrophobicity, which is an essential ingredient for self-cleaning surfaces \cite{kim2014creation}. Further, these morphologies can enable the control of active transport, and expand current applications of focal conic domains in smectics, such as guides for self-assembly of nanoparticles \cite{milette2012reversible}, selective microlens photomasks \cite{kim2010optically}, and building blocks for soft lithography patterning \cite{yoon2007internal}.

In Sec. \ref{sec:inc} we briefly review the model for a smectic-isotropic system of uniform density of Ref. \cite{vitral2019role}. A fully compressible model for smectic-isotropic two phase interface is derived in Sec. \ref{sec:comp}: reversible currents are obtained by imposing zero entropy production in the Second Law of Thermodynamics, while irreversible currents are derived by asserting the Clausius-Duhem inequality holds in case of dissipation. We next specialize this model to a quasi-incompressible case, in which the density of the modulated phase is given constitutively. The density is independent of the pressure but depends on the amplitude of the order parameter. As a consequence, bulk phases are incompressible, but we allow a non-solenoidal velocity in the interfacial region. Compressibility effects are of importance for fluid flow on the surface of the smectic in diffuse-interface treatments, which arise from the difference in density between the two phases. A numerical scheme suitable to study the evolution equation of the order parameter is introduced in Sec. \ref{sec:nm}, which is based on existing schemes for phase field models with varying mobilities. A stability analysis for transverse interfacial perturbations is presented in Sec. \ref{sec:stb}, which is then used to verify the developed numerical code. Finally, in Sec. \ref{sec:flow} we show numerical results concerning flows originating from perturbed smectic layers, and also for layers bent in a focal conic configuration. We discuss the consequences of the varying density field on flow structure, and how curvatures determine interfacial flows through the normal stress balance.


\section{Order parameter model of a smectic-isotropic interface. Incompressible limit}
\label{sec:inc}

We briefly summarize the model of Ref. \cite{vitral2019role} for a smectic-isotropic two phase system. The scalar order parameter $\psi(\mathbf{x},t)$ describes both an isotropic phase with $\psi = 0$, and a smectic phase where $\psi$ is a periodic function of space. This function represents the smectic layered structure, and $\psi$ smoothly changes at the interface between the two phases. A free energy functional of the order parameter is introduced,
\begin{equation}
  {\cal F}_s \;=\; \int \; \frac{1}{2}
            \bigg\{ \epsilon \psi^2 
            + \alpha\left[ \left(\nabla^2+q_0^2 \right)\psi \right]^2        
            - \frac{\beta}{2} \psi^{4} + \frac{\gamma}{3} \psi^{6}\;\bigg\} \,d {\bf x}\; .
  \label{eq:inc-energy}
\end{equation}
where all parameters are constant, including $q_{0}$, the wavenumber of the smectic phase. Relaxational evolution for the order parameter $\psi$ is assumed through minimization of the free energy
\begin{equation}
    \partial_t \psi +\mathbf{v}\cdot\nabla\psi \;=\; 
    -\Gamma\frac{\delta {\cal F}_s}{\delta \psi} \;=\; -\Gamma\mu\;
\label{eq:inc-dynamics}
\end{equation}
where $\mathbf{v}$ is the mass velocity, $\mu$ is the chemical potential conjugate to $\psi$, and $\Gamma$ is a constant mobility (chosen as $\Gamma = 1$ below). Gradient terms in the free energy functional lead to non classical (reversible) stresses of the form, 
\begin{eqnarray}
    \mathbf{T} &=& 
    \frac{\delta {\cal F}_s}{\delta \nabla\mathbf{u}} \;\;=\;\; 
    \nabla\psi\otimes\nabla
    \bigg(\frac{\partial f}{\partial \nabla^2\psi}\bigg)
    -\frac{\partial f}{\partial \nabla^2\psi}\mathbf{D}\psi
    \label{eq:inc-stress}
\end{eqnarray}
where $\mathbf{D}\psi$ is a shorthand for $\partial_i\partial_j\psi$. The stress is defined as the variation of the energy with respect to an independent distortion $\psi(\mathbf{x}) \rightarrow \psi(\mathbf{x}+\mathbf{u})$. Adding dissipative stresses appropriate for an isotropic, Newtonian fluid, lead to the following governing system of equations for an incompressible fluid,
\begin{eqnarray}
    \nabla\cdot\mathbf{v} &=& 0
    \\[2mm]
    \rho\left( \partial_t\mathbf{v} + \mathbf{v}\cdot\nabla\mathbf{v} \right)
    \label{eq:inc-bm}
    &=&
    -\nabla p + \nabla\cdot\mathbf{T} + \eta\nabla^2\mathbf{v}
    \label{eq:inc-blm}
    \\[2mm]
    \partial_t \psi +\mathbf{v}\cdot\nabla\psi &=&
    -\epsilon \,\psi 
    - \alpha \, (\nabla^2+q_0^2)^2 \psi + \beta \,\psi^3 - \gamma \,\psi^5.
    \label{eq:inc-psi}
\end{eqnarray}
In this system, $\rho$ is the constant density, and $\nu$ is the isotropic shear viscosity. It is straightforward to replace the Newtonian viscous dissipation introduced with that of an uniaxial fluid. In the incompressible case, this amounts to considering three independent viscosities. For simplicity, we restrict our analysis here to isotropic viscous dissipation, while noting that the reversible part of the stress does contain the uniaxial symmetry of the smectic phase.

Equation (\ref{eq:inc-psi}) as a model of a smectic-isotropic configuration was investigated in Ref. \cite{vitral2019role}, albeit without advection ($\mathbf{v} = 0$). We analyzed the role of Gaussian curvature on local thermodynamics at the two phase interface (the Gibbs-Thomson equation), and on the evolution of a smectic-isotropic interface, including the effects of local equilibrium thermodynamics from layer alignment with respect to the interface. By examining focal conic instabilities under heat treatment, we showed that conical pyramids of smectic layers could be obtained as observed in experiments \cite{kim2016controlling}, and that their formation could be explained through the interplay between Gaussian curvature, mean curvature and layering alignment. We expect these results to hold qualitatively for a smectic-isotropic fluid interface in terms of the main mechanism of smectic evaporation and condensation. However, in order to fully develop a model that connects to experiments in smectic thin films, we need to account for hydrodynamics and a varying density field between the phases. This way, one can capture the role of surface flows and compressibility effects at the interface, which are relevant not only for the evolution of smectic-isotropic interface, but may also be important for interactions between topological defects in smectics.


\section{Order parameter model of a smectic-isotropic interface. Compressible phases}
\label{sec:comp}

We derive in this section a diffuse interface model for a smectic phase in contact with an isotropic fluid when they have different densities. Earlier work by Brand and Pleiner \cite{brand1980nonlinear} considered a hydrodynamic theory for smectics and other mesophases exhibiting broken symmetries. They introduced an energy density $e$ that depends on mass density, momentum, entropy, and on a variable representing the broken symmetry of the system (e.g. the director $\hat{n}$ and its derivatives in nematics). We use the same methodology but focus on a real variable $\psi$ representing the layering order. Our energy density depends on the Laplacian $\nabla^2\psi$, which leads to layer formation, and accounts for the energy involved in layer distortions. For completeness, Appendix \ref{sec:ap} gives the derivation of the governing equations of our model starting from an Oseen-Frank description and a smectic layer variable, and using the same Coleman-Noll procedure as in this section.

\subsection{Compressible model}

We write the internal energy of the system in terms of the energy per unit mass $e$ and the mass density $\rho$ as
\begin{eqnarray}
    \mathcal{E} &=& \int_\Omega \rho e\, d\mathbf{x} \; .
    \label{eq:energy}
\end{eqnarray}
We first obtain the local form of the internal energy and entropy balances as given by Lowengrub and Truskinovsky for the Cahn-Hilliard equation for a binary fluid \cite{lowengrub1998quasi}. The first law of thermodynamics can be written as,
\begin{equation}
    \frac{d}{dt}(\mathcal{E}+\mathcal{K}) = \mathcal{W} + \mathcal{R} \;,
    \label{eq:ebal}
\end{equation}
where $\mathcal{K}$ is the kinetic energy, $\mathcal{W}$ is the rate of work done on the surface of the system, and $\mathcal{R}$ is the heat transfer rate. They are defined by the following integrals
\begin{eqnarray*}
    \mathcal{K} &=& \int_\Omega \frac{\mathbf{g}^2}{2\rho} d\mathbf{x}, \quad\quad
    \mathcal{R} \;\;=\;\;  \int_\Omega \rho r d\mathbf{x},
    \quad\quad
    \mathcal{W} = \int_{\partial\Omega}\bigg[\mathbf{T}\,\mathbf{n}\cdot\mathbf{v}
    +(\mathbf{t}\cdot\mathbf{n})\dot{\psi}\bigg] dS \; .
\end{eqnarray*}
We use the notation $\dot{(\;)} = \partial_t (\;)+ \mathbf{v}\cdot (\;) $ to denote the material time derivative. Here, $\mathbf{g} = \rho\mathbf{v}$ is the momentum density, $r$ is the rate of heat supplied per unit mass, $\mathbf{T}$ is the stress tensor, $\mathbf{t}$ is the generalized surface force, and $\mathbf{n}$ is the surface normal. The relations we derive in this section are obtained in the absence of thermal radiation, so that we neglect $\mathcal{R}$ for the rest of this section. When deriving the governing equations, we set no-flux boundary conditions: Neumann condition for the order parameter $\psi$ (which forces the smectic planes to be perpendicular to to the domain outer boundary) and zero normal velocity on the boundary, such that
\begin{eqnarray}
    \nabla \psi (\mathbf{x})  \cdot \mathbf{n} = 0, \quad \mathbf{v}(\mathbf{x})\cdot \mathbf{n} = 0, \mathbf{x} \in \partial \Omega.
\label{eq:bc}
\end{eqnarray}

Accounting for the balance of linear momentum $\rho \dot{\mathbf{v}} = \nabla\cdot\mathbf{T}$ and balance of mass $\dot{\rho} + \rho\nabla\cdot\mathbf{v} = 0$, we obtain the local form of the balance of internal energy \cite{gurtin2010mechanics,tadmor2012continuum} as
\begin{eqnarray}
    \rho\dot{e} &=& \mathbf{T}:\nabla \mathbf{v} + \nabla \cdot (\mathbf{t}\,\dot{\psi}) \;.
    \label{eq:ebal-local}
\end{eqnarray}

In order to derive the balance of entropy for the specific internal entropy $s$, we assume $e$ has a dependence not only on $\nabla\psi$, but also on $\nabla^2\psi$, so $e = e(\rho,s,\psi,\nabla \psi,\nabla^2\psi)$. This dependence on $\nabla^2\psi$ does not appear for binary systems, but is fundamental to model the smectic phase. Hence, by the chain rule
\begin{equation*}
    \dot{e} = \frac{\partial e}{\partial \rho}\dot{\rho}
        +\frac{\partial e}{\partial s}\dot{s} 
        + \frac{\partial e}{\partial \psi}\dot{\psi}
        + \frac{\partial e}{\partial \nabla\psi}\cdot\dot{\overline{\nabla\psi}}
        + \frac{\partial e}{\partial \nabla^2\psi}\dot{\overline{\nabla^2\psi}} \; .
\end{equation*}
where the overbar notation denotes the material time derivative of the entire term bellow the bar. Given that the temperature $\theta = \partial e/ \partial s$, we rewrite Eq. (\ref{eq:ebal-local}) as a local balance of entropy
\begin{eqnarray}
    \nonumber
    &&\rho \theta \dot{s} \;\;= 
    \\[2mm] \nonumber 
        && \bigg\{ \mathbf{T} + \rho^2\frac{\partial e}{\partial \rho}\mathbf{I}
        + \rho \nabla\psi \otimes \frac{\partial e}{\partial\nabla\psi}
        -\nabla\psi\otimes\nabla
        \bigg(\rho\frac{\partial e}{\partial \nabla^2\psi}\bigg)
        +\rho\frac{\partial e}{\partial \nabla^2\psi}\mathbf{D}\psi
    \bigg\} : \nabla \mathbf{v} \\[2mm]
    && +\bigg[ \mathbf{t} -\rho \frac{\partial e}{\partial\nabla\psi} + \nabla\left(\rho\frac{\partial e}{\partial \nabla^2 \psi}\right
        )\bigg] \cdot \nabla \dot{\psi} - \bigg[\rho\frac{\partial e}{\partial \psi}-\nabla\cdot\mathbf{t}\bigg]\dot{\psi} \; ,
\label{eq:sbal}
\end{eqnarray}
where $\mathbf{D}$ stands for $\partial_i\partial_j$, so that $\mathbf{D}\psi$ is a second order tensor. In deriving the previous expression, the boundary conditions from Eq. (\ref{eq:bc}) allow us to write
\begin{flalign*}
    &\rho \frac{\partial e}{\partial \nabla^2\psi}\dot{\overline{\nabla^2\psi}} \;\;=\;\;
    \rho \frac{\partial e}{\partial \nabla^2\psi}\nabla^2\dot{\psi} 
    - \rho \frac{\partial e}{\partial \nabla^2\psi}\nabla^2\mathbf{v}\cdot\nabla\psi
    -2\rho \frac{\partial e}{\partial \nabla^2\psi}\nabla\mathbf{v} : \mathbf{D}\psi &
   \\[2mm] &\hspace{10mm} =\quad 
    -\nabla\bigg(\rho\frac{\partial e}{\partial\nabla^2\psi}\bigg)\cdot\nabla\dot{\psi}
    + \bigg[ \nabla\psi\otimes\nabla\bigg(\rho \frac{\partial e}{\partial\nabla^2\psi}\bigg)
    - \rho \frac{\partial e}{\partial\nabla^2\psi} \mathbf{D}\psi \bigg] : \nabla\mathbf{v} \; ,&
\end{flalign*}
and also
\begin{eqnarray*}
 \rho \frac{\partial e}{\partial \nabla\psi}\cdot\dot{\overline{\nabla\psi}} &=& 
 \rho\frac{\partial e}{\partial \nabla\psi}\cdot \nabla\dot{\psi} 
 - \rho \nabla\psi \otimes \frac{\partial e}{\partial\nabla\psi} : \nabla\mathbf{v} \; .
\end{eqnarray*}
The terms in square brackets proportional to $\dot{\psi}$ and $\nabla\dot{\psi}$ in Eq. (\ref{eq:sbal}) are both related to variations of $\psi$ and can be grouped together. By using the boundary conditions, we write
\begin{eqnarray}
    \nonumber
    &&\rho \theta \dot{s} \;\;= 
    \\[2mm] \nonumber
    && \bigg\{ \mathbf{T} + \rho^2\frac{\partial e}{\partial \rho}\mathbf{I}
        +\rho \nabla\psi \otimes \frac{\partial e}{\partial\nabla\psi}
        -\nabla\psi\otimes\nabla
        \bigg(\rho\frac{\partial e}{\partial \nabla^2\psi}\bigg)
        +\rho\frac{\partial e}{\partial \nabla^2\psi}\mathbf{D}\psi
    \bigg\} : \nabla \mathbf{v} \\[2mm]
    && +\bigg[-\rho\frac{\partial e}{\partial \psi}
    +\nabla\cdot\bigg(\rho \frac{\partial e}{\partial\nabla\psi}\bigg)
    -\nabla^2\bigg(\rho\frac{\partial e}{\partial \nabla^2 \psi}\bigg) \bigg] \dot{\psi}
    \; .
\label{eq:sbal2}
\end{eqnarray}

In order to obtain the required constitutive relations, we use the Coleman-Noll procedure, which defines necessary conditions for them by imposing a strict requirement on the entropy production. Based on the Clausius-Duhem inequality, the condition for the specific internal entropy $\dot{s} \geq 0$ implies that Eq. (\ref{eq:sbal2}) must be satisfied for every admissible thermomechanical process. Hence, by splitting the stress into reversible and dissipative parts, $\mathbf{T} = \mathbf{T}^R + \mathbf{T}^D$, we can derive the reversible parts from Eq. (\ref{eq:sbal2}) in the limit of zero entropy production, while dissipative parts are obtained by enforcing positive entropy production.

For deriving $\mathbf{T}^R$, which is a reversible current for the balance of linear momentum, we set the terms in brackets associated with the rates $\nabla\mathbf{v}$ equal to zero, so that
\begin{eqnarray}
    \mathbf{T}^R &=& -\rho^2\frac{\partial e}{\partial \rho}\mathbf{I}
        - \rho \nabla\psi \otimes \frac{\partial e}{\partial\nabla\psi}
        +\nabla\psi\otimes\nabla
        \bigg(\rho\frac{\partial e}{\partial \nabla^2\psi}\bigg)
        -\rho\frac{\partial e}{\partial \nabla^2\psi}\mathbf{D}\psi \; .
        \label{eq:rev-stress}
\end{eqnarray}

The expression in square brackets multiplying $\dot{\psi}$ is the thermodynamic conjugate to $\psi$, $\mu = \delta \mathcal{E} / \delta \psi$. That is,
\begin{eqnarray}
    \mu &=& \rho\frac{\partial e}{\partial \psi}
    -\nabla\cdot\bigg(\rho \frac{\partial e}{\partial\nabla\psi}\bigg)
    +\nabla^2\bigg(\rho\frac{\partial e}{\partial \nabla^2 \psi}\bigg) \; .
    \label{eq:chemicalp}
\end{eqnarray}
Since $\dot{s} = 0$ for reversible motions, and $\mu$ is arbitrary, we must have $\dot{\psi} = 0$.  The order parameter $\psi$ is a slowly relaxing variable, which is not associated with any conservation law. Hence, its dynamic equation is of the form
\begin{eqnarray}
    \partial_t \psi + \mathbf{v}\cdot\nabla\psi + Z &=& 0
    \label{eq:dyn-psi}
\end{eqnarray}
where $Z$ is a quasi-current (that is, its surface integral is not a flux), which can be decomposed into $Z = Z^R + Z^D$. Since in the reversible limit $\dot{\psi} = 0$, this implies that $Z^R = 0$, and we are only left with the dissipative part $Z^D$. 

To obtain the form of the irreversible currents, we need to impose the condition $\dot{s} > 0$ to Eq. (\ref{eq:sbal2}) (one can also derive these functions from derivatives of a generalized function with respect to thermodynamic forces \cite{brand2001macroscopic,pleiner1996hydrodynamics}). This implies that $\dot{\psi}$ must be proportional to the negative of the chemical potential times a constant $\Gamma$, so that $Z^D$ has the form
\begin{eqnarray}
    Z^D &=& \Gamma \mu \; .
    \label{eq:dis-z}
\end{eqnarray}
Physically, the dissipative contribution to $\dot{\psi}$ is a permeation mode \cite{degennes1995physics}; it is nonzero when there is mass transport relative to the smectic layers.

We finally introduce the dissipative contribution to the stress. When $\dot{s} > 0$, only $\mathbf{T}^D$ remains inside the curly brackets contracted with $\nabla\mathbf{v}$ in Eq. (\ref{eq:sbal2}), so that to enforce positive entropy production we require
\begin{eqnarray}
    \mathbf{T}^D &=& {\boldsymbol \eta}:\nabla \mathbf{v} \; .
    \label{eq:dis-stress}
\end{eqnarray}
For an uniaxial phase with an optical axis $\mathbf{n}$ (normal direction to the smectic layers), by defining the rate of deformation tensor as $\mathbf{E} = (\nabla\mathbf{v}+\nabla\mathbf{v}^\mathrm{T})/2$, this viscous term presents five independent viscosities $\alpha_{i(j)}$ \cite{degennes1995physics,martin1972unified}
\begin{eqnarray*}
  \eta_{ijkl}\partial_lv_k
  &=& \alpha_0\delta_{ij}E_{kk} + \alpha_1\delta_{in}\delta_{jn}E_{nn}
      +\alpha_4E_{ij}
  \\ &&
        +\alpha_{56}(\delta_{in}E_{nj}+\delta_{jn}E_{ni})
        +\alpha_7(\delta_{in}\delta_{jn}E_{kk}+\delta_{ij}E_{nn}) \;.
\end{eqnarray*}

For simplicity, we will restrict our study to the case of a Newtonian fluid for both phases, although the extension to a uniaxial fluid is simple. Therefore, instead of working with the full fourth order viscosity tensor ${\boldsymbol \eta}$, we consider only the viscosity coefficient $\eta$, and the second coefficient of viscosity $\lambda$, which are also assumed to be the same for both phases. Because we account for compressibility effects on the interface, the velocity is non-solenoidal, which adds a contribution function of $\nabla\cdot\mathbf{v}$ to the viscous stress
\begin{eqnarray}
    \mathbf{T}^D &=& \eta (\nabla\mathbf{v}+\nabla\mathbf{v}^\mathrm{T}) +
    \lambda(\nabla\cdot\mathbf{v})\mathbf{I} \; .
    \label{eq:dis-stress2}
\end{eqnarray}
While out of equilibrium the two viscosities are generally independent, in our numerical investigations we follow Stokes' hypothesis and set the bulk viscosity to zero. That is, we set trace ($\mathbf{T}^D) = 0$, which gives $\lambda = -\frac{2}{3} \eta$. The choice of $\lambda$ has an important impact on the fully compressible model, as it controls the magnitude of the longitudinal part of the flow. When one accounts for the viscous term with all five viscosity coefficients, the longitudinal flow also depends on derivatives of the velocity field with respect to the direction of $\nabla\psi$, damping oscillations of the flow along the optical axis. A full analysis would require extensive numerical study on the role of each viscosity coefficient, as well as evaluation of the ratios among these coefficients to best connect to empirical data.

The equations governing the evolution of the quasi incompressible system now read
\begin{eqnarray}
    \dot{\rho} &=& 
    -\rho\nabla\cdot\mathbf{v} \; ,
    \label{eq:com-bm}
    \\[2mm]
    \rho\dot{\mathbf{v}} &=&
    \nabla\cdot\Big(\mathbf{T}^R + \mathbf{T}^D\Big) \; ,
    \label{eq:com-blm}
    \\[2mm]
    \dot{\psi} &=& -\Gamma \mu \; ,
    \label{eq:com-psi}
\end{eqnarray}
with $\mathbf{T}^{R}$ defined in Eq. (\ref{eq:rev-stress}), $\mathbf{T}^{D}$ in Eq. (\ref{eq:dis-stress2}) and $\mu$ in Eq. (\ref{eq:chemicalp}). Boundary conditions on the outer boundaries are specified by Eqs. (\ref{eq:bc}). 

\subsection{Quasi-incompressible model}

We next assume that the density does not depend on pressure in the bulk phases (quasi incompressible assumption), but depends constitutively on $\psi$, so that $\rho = \rho(\psi)$ in the two phase system. Due to the modulated nature of $\psi$, the choice of constitutive relation is not as straightforward as in the Cahn-Hilliard model of a binary mixture in which there is a transition between two regions of uniform composition. We write in the present case $\rho$ as a function only of the slowly varying envelope of the order parameter, $A(\mathbf{x})$, as defined in Ref. \cite{vitral2019role},
\begin{eqnarray}
    \rho[A(\mathbf{x})] &=& \kappa A(\mathbf{x}) + \rho_0 \; ,
    \label{eq:density}
\end{eqnarray}
where $\kappa$ is a constant that controls the density ratio between the bulk smectic and isotropic phases, and $\rho_{0}$ is the density of the isotropic phase where $A = 0$. In practice, we compute the amplitude by using $A = (\psi^2+q_0^{-2}|\nabla\psi|^2)^{1/2}$. For the form of the energy that we introduce below, we have numerically confirmed that for smectic layers that are not severely distorted this expression accurately captures the amplitude of $\psi$. 

While both bulk fluids are incompressible, the velocity field becomes non-solenoidal at the interface. From the balance of mass in Eq. (\ref{eq:com-bm}), we find that
\begin{eqnarray}
    \nabla\cdot\mathbf{v} &=& 
    -\frac{\partial\rho}{\partial A}\frac{\dot{A}}{\rho} 
    \;\;=\;\; -\kappa\frac{\dot{A}}{\rho}\; .
    \label{eq:divv}
\end{eqnarray}
We note that Eq. (\ref{eq:divv}) is similar to that used for the Cahn-Hilliard model of a quasi-incompressible binary fluid, which becomes more clear by expressing it in terms of $\psi$. The material time derivative of $A$ is connected to permeation, that is mass motion relative to smectic planes, so that the divergence of the velocity in a quasi-incompressible diffusive-interface model is linked to the order parameter chemical potential, as discussed in Refs. \cite{lowengrub1998quasi,shokrpour2018diffuse}. From $\rho = \rho (\psi)$, one can also write
\begin{eqnarray}
    \nabla\cdot\mathbf{v} &=& 
    -\frac{\partial \rho}{\partial \psi}\frac{\dot{\psi}}{\rho} 
    \;\;=\;\; \frac{\partial\rho}{\partial\psi}\frac{\Gamma\mu}{\rho} \;.
    \label{eq:divv-psi}
\end{eqnarray}

Finally, we make explicit the dependence of the chemical potential on pressure by decomposing the velocity gradient $\nabla\mathbf{v} = \mathbf{S_v} + \frac{1}{3} (\nabla\cdot\mathbf{v}) \mathbf{I}$,  where $\mathbf{S_v}$ is its deviatoric part. We can rewrite the local balance of entropy from Eq. (\ref{eq:sbal2}), so that the stress contracts with the deviatoric tensor $\mathbf{S_v}$. Since $\mathbf{I}:\mathbf{S_v} = 0$, any scalar multiplying the identity in the stress satisfies the Clausius-Duhem inequality. Therefore, we introduce the pressure $p$, which is not uniquely defined, and write the reversible part of the stress as
\begin{eqnarray}
    \mathbf{T}^R &=& -p\mathbf{I}
    - \rho \nabla\psi \otimes \frac{\partial e}{\partial\nabla\psi}
    +\nabla\psi\otimes\nabla
    \bigg(\rho\frac{\partial e}{\partial \nabla^2\psi}\bigg)
    -\rho\frac{\partial e}{\partial \nabla^2\psi}\mathbf{D}\psi \; .
    \label{eq:rev-stress2}
\end{eqnarray}
When the identity contracts with the stress terms inside the curly brackets from Eq. (\ref{eq:sbal2}), we get exactly $3p$ from the resulting trace. Using this result and substituting $\nabla\cdot\mathbf{v}$ from Eq. (\ref{eq:divv-psi}) in the local balance of entropy, we write
\begin{eqnarray}
    \nonumber
    \rho \theta \dot{s} &=&
    \bigg\{ \mathbf{T}
    +\rho \nabla\psi \otimes \frac{\partial e}{\partial\nabla\psi}
    -\nabla\psi\otimes\nabla
    \bigg(\rho\frac{\partial e}{\partial \nabla^2\psi}\bigg)
    +\rho\frac{\partial e}{\partial \nabla^2\psi}\mathbf{D}\psi
    \bigg\} : \mathbf{S_v} \\
     &&+\bigg[
    p\, \rho^{-1} \frac{\partial\rho}{\partial\psi}
    -\rho\frac{\partial e}{\partial \psi}
    +\nabla\cdot\bigg(\rho \frac{\partial e}{\partial\nabla\psi}\bigg)
    -\nabla^2\bigg(\rho\frac{\partial e}{\partial \nabla^2 \psi}\bigg) \bigg] \dot{\psi} \; ,
    \label{eq:sbal3}
\end{eqnarray}
Therefore, the order parameter chemical potential per unit volume now exhibits an explicit dependence on the kinematic pressure, that is
\begin{eqnarray}
    \mu &=&
    -p\,\rho^{-1}\frac{\partial\rho}{\partial\psi}
    +\rho\frac{\partial e}{\partial \psi}
    -\nabla\cdot\bigg(\rho \frac{\partial e}{\partial\nabla\psi}\bigg)
    +\nabla^2\bigg(\rho\frac{\partial e}{\partial \nabla^2 \psi}\bigg) \;.
    \label{eq:quasi-chemicalp}
\end{eqnarray}
The governing equations are given by Eqs. (\ref{eq:com-blm}) and (\ref{eq:com-psi}), with the definitions for the chemical potential and reversible stress as given by Eqs. (\ref{eq:quasi-chemicalp}) and (\ref{eq:rev-stress2}) respectively. 

\subsection{Choice of energy functional}

In order to study a system comprising a smectic and an isotropic phase which can achieve coexistence, we choose \cite{vitral2019role},
\begin{eqnarray}
    e(\psi,\nabla^2\psi) &=& \frac{1}{2}\bigg\{ \epsilon \psi^2 
    + \alpha\left[ \left(\nabla^2+q_0^2 \right)\psi \right]^2            
    - \frac{\beta}{2} \psi^{4} + \frac{\gamma}{3} \psi^{6}\;\bigg\}.
    \label{eq:energy-density}
\end{eqnarray}
The coefficients $\alpha$, $\beta$ and $\gamma$ are three constant, positive parameters, and $\epsilon$ is a bifurcation parameter that describes the distance away from the smectic-isotropic transition. The values of the constants $\beta$ and $\gamma$ are chosen to give a triple well energy function with minima representing smectic and isotropic phases \cite{sakaguchi1996stable}. Coexistence occurs at $\epsilon_{c} = 27 \beta^{2}/160 \gamma$, when both phases present the same energy density. For $\epsilon > \epsilon_{c}$, $\psi = 0$ becomes the equilibrium phase, whereas for $\epsilon < \epsilon_{c}$, a modulated phase $\psi \approx \frac{1}{2} (A\,e^{i{\bf q}\cdot{\bf x}} + c.c.)$ is stable. Here $| \mathbf{q} | \approx q_{0}$, with $\mathbf{q}$ along an arbitrary direction. 

The chemical potential from Eq. (\ref{eq:quasi-chemicalp}) is now,
\begin{eqnarray}
    \mu &=& -p\,\rho^{-1}\frac{\partial\rho}{\partial\psi} +\rho\Big[\epsilon\psi +\alpha q_0^2(\nabla^2+q_0^2)\psi-\beta\psi^3+\gamma\psi^5\Big] 
    + \alpha\nabla^2\Big[\rho(\nabla^2+q_0^2)\psi\Big]\;.
    \label{eq:quasi-chemicalp2}
\end{eqnarray}
One important remark about computing $\mu$ is that $\rho$ in Eq. (\ref{eq:density}) is given as a function of the amplitude $A$, so we do not actually have an expression for $\rho(\psi)$. By the chain rule, a simple calculation of $\partial\rho / \partial\psi$ from the way we obtain $A$ from a $\psi$ would give $\kappa\psi/A$. Since we set $\rho$ as constant in the bulk of the two phases, its derivative with respect to $\psi$ should only be nonzero at the interface. Due to this, we interpret $\psi$ that shows in the previous derivative as the average of the order parameter, $\langle\psi\rangle$, computed over a unit cell defined by the wavelength (which is zero outside the interface). For parallel computations, it becomes costly to perform such averaging after every iteration, so that an alternative is to approximate $\langle\psi\rangle/A$ by $|\nabla A|/A$, or similarly $|\nabla\rho|/\rho$.

The balance of linear momentum for the choice of energy given by Eq. (\ref{eq:energy-density}) is
\begin{eqnarray}
    \rho\dot{\mathbf{v}} &=& -\nabla p 
    + \nabla^2\bigg(\rho\frac{\partial e}{\partial\nabla^2\psi}\bigg)\nabla\psi
    -\rho\frac{\partial e}{\partial\nabla^2\psi}\nabla^2\nabla\psi 
    + \eta\nabla^2\mathbf{v} 
    + (\lambda+\eta)\nabla(\nabla\cdot\mathbf{v}) \;.
    \label{eq:quasi-ns}
\end{eqnarray}
By focusing only on overdamped or Stokes flow, we further assume that the fluid velocity everywhere satisfies 
\begin{eqnarray}
    \nonumber
    0 &=& -\nabla p + \alpha\nabla^2\Big[\rho(\nabla^2+q_0^2)\psi\Big]\nabla\psi
    -\alpha\rho(\nabla^2+q_0^2)\psi\nabla^2\nabla\psi
    + \eta\nabla^2\mathbf{v} 
    + (\lambda+\eta)\nabla(\nabla\cdot\mathbf{v}) \; .
    \label{eq:quasi-stokes}
\end{eqnarray}
Taking the divergence of Eq. (\ref{eq:quasi-stokes}), once $\psi$ is known, $p$ can be immediately obtained through a modified pressure Poisson equation. 

\subsection{Governing equations in dimensionless form}

By using the constitutive law Eq. (\ref{eq:density}) we summarize here the complete set of governing equations for the smectic-isotropic fluid system,
\begin{eqnarray}
    \nabla\cdot\mathbf{v} &=& 
    -\kappa \frac{\dot{A}}{\rho}\; ,
    \label{eq:quasi-divv}
    \\[4mm]
    \nonumber
    0 &=& -\nabla p + \alpha\nabla^2\Big[\rho(\nabla^2+q_0^2)\psi\Big]\nabla\psi
    -\alpha\rho(\nabla^2+q_0^2)\psi\nabla^2\nabla\psi
    \\[2mm] &&
    + \eta\nabla^2\mathbf{v} 
    + (\lambda+\eta)\nabla(\nabla\cdot\mathbf{v}) \; ,
    \label{eq:quasi-blm}
    \\[4mm]
    \dot{\psi} &=& -\Gamma \mu \; ,
    \label{eq:quasi-psi}
    \\[4mm]
     \rho &=& \kappa A + \rho_0 \; .
    \label{eq:quasi-density}
\end{eqnarray}

To introduce dimensionless variables, let U and L represent characteristic scales for the velocity and length, and $\tilde{\rho}$, $\tilde{\psi}$ and $\tilde{\mu}$ represent typical values for $\rho$, $\psi$ and $\mu$ in the modulated phase. Then, we introduce the dimensionless variables $\mathbf{v}^* = \mathbf{v}/U$, $\mathbf{x}^* = \mathbf{x}/L$, $t^* = U t/L$,  $\rho^* = \rho/\tilde{\rho}$, $\psi^* = \psi/\tilde{\psi}$ and $\mu^* = \mu/\tilde{\mu}$.  The resulting equations have the same form as Eqs. (\ref{eq:quasi-divv})-(\ref{eq:quasi-density}), replacing constants and variables by dimensionless constants and variables. The dimensionless constants one finds are $\tilde{\kappa} = \kappa / \tilde{\rho}$, $\tilde{\Gamma} = \Gamma L\tilde{\mu}/\tilde{\psi}U$, $\tilde{\eta} = \eta U L^3 / \tilde{\rho}\tilde{\psi}^2$ and $\tilde{\lambda} = \lambda U L^3 / \tilde{\rho}\tilde{\psi}^2$, where the last two are proportional to the capillary number. In the following discussion, we use the non-dimensional set of governing equations, dropping the tilde from constants and star from variables.


\section{Numerical Method}
\label{sec:nm}

We solve Eqs. (\ref{eq:quasi-divv})-(\ref{eq:quasi-density}) numerically, with boundary conditions specified in Eq. (\ref{eq:bc}), by using a pseudo-spectral method, in which linear and gradient terms are computed in Fourier space and nonlinear terms in real space. Space discretization depends on $n_w$, the number of points per base wavelength, and is given by $\Delta x = 2\pi/(n_w q_0)$. The appropriate choice of time step will be later analyzed in the context of the scheme stability. We have developed custom C++ codes based on the parallel FFTW library and the standard MPI passing interface for parallelization. In order to accommodate the boundary conditions, we use both the Discrete Cosine Transform of ($\psi$, $\rho$) and the Discrete Sine Transform of ($\nabla\psi$, $\mathbf{v}$). The source codes containing the implementation of this model (\textit{smaiso-quasi}) can be found in Ref. \cite{smaiso-quasi}, and the codes for the simpler uniform density model (\textit{smaiso-uniform}) described in Sec. \ref{sec:inc} are found in Ref. \cite{smaiso-uniform}.

In our previous work on the smectic-isotropic (constant density) problem \cite{vitral2019role}, we integrated the dynamic equation for $\psi$ in time employing a Crank-Nicolson algorithm for the linear part of the equation, and a second order Adams-Bashforth method for the nonlinear terms. However, we cannot deal with Eq. (\ref{eq:quasi-psi}) in the same way (splitting it into linear and nonlinear parts), as now the right hand side is multiplied by a varying density. Therefore, we rewrite Eq. (\ref{eq:quasi-psi}) as $\partial_t\psi = \Gamma(\rho L \psi + N)$ with
\begin{eqnarray}
    L &=& -\Big[\epsilon + (\nabla^2+q_0^2)^2\Big]
    \label{eq:linear}
    \\[2mm]
    N &=& \frac{p}{\rho}\frac{\partial\rho}{\partial\psi}-2\alpha\nabla\rho\cdot(\nabla^2+q_0^2)\nabla\psi
    -\alpha\nabla^2\rho\,(\nabla^2+q_0^2)\psi+\beta\psi^3-\gamma\psi^5
    -\Gamma^{-1}\mathbf{v}\cdot\nabla\psi
    \label{eq:nlinear}
\end{eqnarray}
where $L$ is a linear operator, and $N$ is a collection of nonlinear terms. Note that $\Gamma\rho$ plays the role of a spatially varying mobility (this is why we cannot treat this equation as in Ref. \cite{vitral2019role}). We follow a scheme already introduced for phase-field models with variable mobility \cite{zhu1999coarsening,badalassi2003computation}: We split the density as $\rho \rightarrow \rho_m +(\rho-\rho_m)$, where $\rho_m = \frac{1}{2}(\rho_s + \rho_0)$. Here, $\rho_s$ is the density of the smectic bulk, which can be obtained from the system parameters by $\rho_s = \kappa A_s + \rho_0$, where $A_s = 2 A_0$ is the amplitude of the sinusoidal phase, with $A_0$ given by Eq. (\ref{eq:a0}). The idea behind the split is that the term associated to $\rho_m$ can be treated implicitly, and $(\rho-\rho_m)$ explicitly, with a choice of $\rho_m$ that satisfies $|\rho -\rho_m| \leq \rho_m$. 

In Fourier space, we use a second order discretization in time, and compute $\psi$ at time $n+1$ by
\begin{eqnarray}
    \frac{\frac{3}{2}\psi_k^{n+1}-2\psi_k^n+\frac{1}{2}\psi_k^{n-1}}{\Delta t} 
    &=& \Gamma\Big[\rho_m L\psi_k^{n+1}-\rho_m L\psi_k^n 
    + (\rho^n L \psi^n + N^n)_k\Big]
    \label{eq:method}
\end{eqnarray}
The term $\rho^n L \psi^n$ is nonlinear, so we include it in the definition of $N$. Instead of solely accounting for the nonlinear terms $N$ at time $n$, we treat $N$ with a second order multistep Adams-Bashforth scheme. In frequency space, $\psi_k$ for the new time is then obtained by
\begin{eqnarray}
    (3/2-\Delta t\, \Gamma\rho_m L )\psi_k^{n+1}
    &=&  (2-\Delta t\, \Gamma\rho_m L)\psi_k^{n}-\frac{1}{2}\psi_k^{n-1}
    +\frac{\Delta t\, \Gamma}{2}(3N_k^n-N_k^{n-1}) \; .
    \label{eq:method2}
\end{eqnarray}

Overall, our model -as well as the physical system- is only concerned with a slowly varying density, on the scale of variations of the envelope $A = (\psi^2+q_0^{-2} |\nabla\psi|^2)^{1/2}$, but not changing on the scale of the smectic layers, $1/q_{0}$.  While this approximation for $A$ gives us an adequate approximation for the amplitude of $\psi$ in regions where the smectic layers are well formed and only weakly distorted, it becomes noisier on the interface and also in regions where layers are highly distorted or break up. Therefore in our numerical calculations we smooth the computed amplitude with a Gaussian filter in Fourier space, given by the operator $F_{\zeta} = \textrm{exp}(-\zeta^2 q^2/2)$, where $q$ is the wavenumber and $\zeta$ the filtering radius, chosen as $1/q_0$. For large density ratios, $(\rho_s-\rho_0)/\rho_0 > 5$, we also observe for numerical instabilities originating from terms containing gradients of $\rho$ in Eqs. (\ref{eq:quasi-chemicalp2}) and (\ref{eq:quasi-blm}), due to fast oscillatory terms that should be compensated by an oscillatory pressure. Therefore, while in such cases we use a spatially varying density in the numerical integration, we neglect higher order terms in terms $\nabla\rho$ and $\nabla^2\rho$ from Eqs. (\ref{eq:quasi-blm}) and (\ref{eq:quasi-psi}).


\section{Stability analysis}
\label{sec:stb}

\begin{figure}[ht]
    \centering
    \includegraphics[width=6cm, height=5cm]{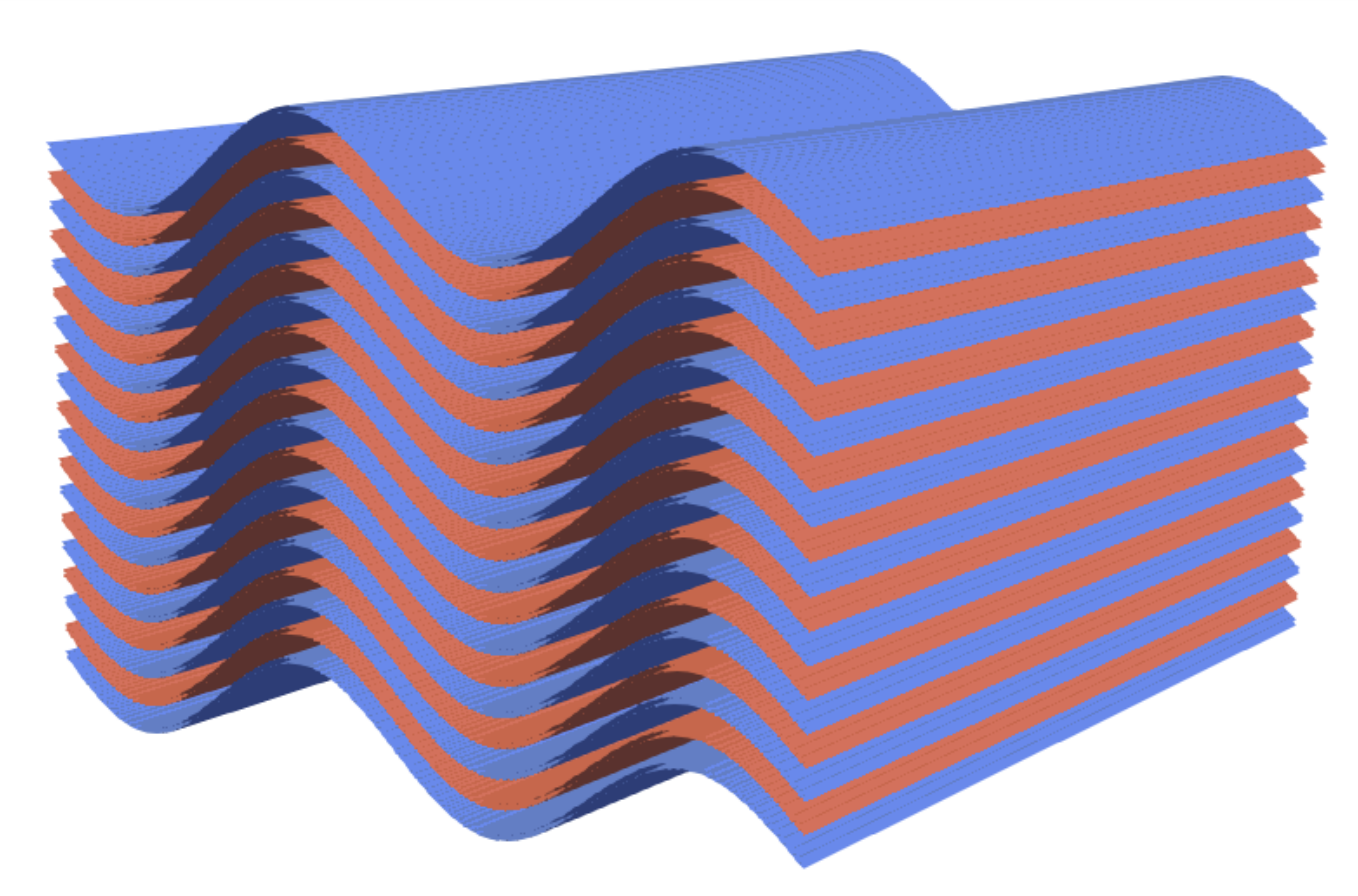}
    \caption{Stack of smectic layers perturbed in the transverse direction. The layers are extracted from a configuration of the order parameter $\psi$, which oscillates between $\psi_{max}$ and $\psi_{min}$ values, defining the red (positive) and blue (negative) layering shown in the figure.}
    \label{fig:smectic}
\end{figure}

In order to elucidate the role of hydrodynamics on interfacial motion, as well as to validate the numerical algorithm, we first address the linear stability of a stack of smectic layers as shown in Fig. \ref{fig:smectic}, and derive the dispersion relation for transverse perturbations of the smectic layers as a function of the distortion wavelength. 
Consider a reference configuration comprising a set of parallel smectic planes that span the whole domain, aligned along a reference wave vector $\mathbf{q}$. The base solution is $\psi_0 = A_0 e^{i\mathbf{q}\cdot\mathbf{x}}+c.c.$, and homogeneous density $\rho = \rho_s$. We introduce a perturbation of wave vector $\mathbf{Q}$ while leaving the density constant,
\begin{eqnarray}
    \psi &=& A_0 e^{i\mathbf{q}\cdot\mathbf{x}}
    +A_1 e^{i(\mathbf{q}+\mathbf{Q})\cdot\mathbf{x}}
    + A_2 e^{i(\mathbf{q}-\mathbf{Q})\cdot\mathbf{x}} + c.c. \; ,
    \label{eq:psi-dist}
\end{eqnarray}
where $A_1, A_2 \ll A_0$ are small amplitudes. Since the density is constant, the order parameter equation reduces to
\begin{eqnarray}
    \partial_t\psi+\mathbf{v}\cdot\nabla\psi &=& -\Gamma\mu \;\;=\;\; \Gamma\rho_s\Big[-\epsilon\psi -\alpha q_0^2(\nabla^2+q_0^2)^2\psi+\beta\psi^3-\gamma\psi^5\Big] \;.
    \label{eq:psi-stb}
\end{eqnarray}
The mobility $\Gamma = 1$ in all our simulations. Define
$$ 
l_0 = |\mathbf{q}|^2 - q_0^2,\quad l_1 = |\mathbf{q}+\mathbf{Q}|^2 - q_0^2, \quad  l_2 = |\mathbf{q}-\mathbf{Q}|^2 - q_0^2 \; .
$$
Then, by keeping only modes exp($\pm i\mathbf{q}\cdot\mathbf{x}$) and exp($\pm i(\mathbf{q}\pm\mathbf{Q}\cdot\mathbf{x})$) when expanding $\mu$ in terms of the perturbation, we find
\begin{eqnarray*}
    \mu \rho_s^{-1} &=& 
    M_0 e^{i\mathbf{q}\cdot\mathbf{x}}
    +M_1e^{i(\mathbf{q}+\mathbf{Q})\cdot\mathbf{x}}
    +M_2e^{i(\mathbf{q}-\mathbf{Q})\cdot\mathbf{x}} + c.c.
\end{eqnarray*}
with
\begin{eqnarray*}
M_0 &=& \epsilon A_0 + l_0^2 A_0 - 3\beta |A_0|^2 A_0 + 10\gamma |A_0|^4 A_0 \;,
\\[2mm]
M_1 &=& \epsilon A_1 + l_1^2 A_1 - 6\beta |A_0|^2 A_1 - 3\beta A_0^2 A_2^* \;,
+30\gamma |A_0|^4 A_1 + 20 \gamma |A_0|^2 A_0^2 A_2^*
\\[2mm]
M_2 &=& \epsilon A_2 + l_2^2 A_2 - 6\beta |A_0|^2 A_2 - 3\beta A_0^2 A_1^*
+30\gamma |A_0|^4 A_2 + 20 \gamma |A_0|^2 A_0^2 A_1^* \;.
\end{eqnarray*}

Since the density is uniform in the smectic layer, the velocity field is solenoidal. Also, the pressure in Eq. (\ref{eq:quasi-blm}) can be redefined so that the momentum balance equation can be written in terms of a forcing term $\mathbf{f} = \mu\nabla\psi$,
\begin{eqnarray}
0 &=& -\nabla \bar{p} +\mathbf{f} + \eta \nabla^2\mathbf{v} \;,
\label{eq:blm-stb}
\\[2mm]
\bar{p} &=& p + \frac{\alpha\rho_s}{2}[(\nabla^2+q_0^2)\psi]^2
+\frac{\rho_s \epsilon}{2}\psi^2-\frac{\rho_s\beta}{4}\psi^4
+\frac{\rho_s\gamma}{6}\psi^6 \;.
\end{eqnarray}
That is, $\bar{p} = p + \rho_s e$. Since for planar smectic layers the chemical potential $\mu$ is zero in equilibrium, the velocity field $\mathbf{v}_0$ for the base $\psi$ solution is also exactly zero.

In order to obtain an expression for the perturbed flow velocity, we set the base state of the smectic layers to be aligned along $z$, $\mathbf{q} = q_0\hat{z}$, so that $\mathbf{Q}$ is orthogonal to $z$. By applying the Fourier transform, we obtain the following terms from $\mathbf{f}$ for frequencies $Q$ and $2q$, in Fourier space,
\begin{eqnarray*}
{\bf f}_\mathbf{Q} &=& \rho_s\Big[ -i M_0 (\mathbf{q}-\mathbf{Q})A_2^* +i M_2^* \mathbf{q} A_0
- i M_1 \mathbf{q}A_0^* + i M_0^*(\mathbf{q}+\mathbf{Q})A_1 \Big]
\\[2mm]
{\bf f}_{\mathbf{2q}} &=& -i \rho_s A_0 \mathbf{q}_0
\end{eqnarray*}
The remaining modes that are required for the leading order expansion of the order parameter equation are given by ${\bf f}_{-\mathbf{Q}}^* = {\bf f}_\mathbf{Q}$ and ${\bf f}_{-2\mathbf{q}}^* = {\bf f}_{2\mathbf{q}}$. By taking the divergence of Eq. (\ref{eq:blm-stb}), we find a pressure Poisson equation, which allows us to calculate the pressure in terms of the frequency $\mathbf{k}$ as
\begin{eqnarray*}
p_k &=& \frac{i\,\mathbf{k}\cdot {\bf f}_\mathbf{k}}{|\mathbf{k}|^2} \;.
\end{eqnarray*}
Then, by substituting the pressure into Eq. (\ref{eq:blm-stb}), we obtain an expression for the flow velocity in terms of the Fourier modes
\begin{eqnarray}
\mathbf{v} &=& \sum_{\mathbf{k}=\pm\mathbf{Q}}\frac{1}{\eta|\mathbf{k}|^2}\bigg(\mathbf{I}-\frac{\mathbf{k}\otimes\mathbf{k}}{|\mathbf{k}|^2}\bigg)\mathbf{f}_\mathbf{k}\, e^{i\mathbf{k}\cdot\mathbf{x}} \;,
\label{eq:vel-stb}
\end{eqnarray}
for which the longitudinal modes, $\mathbf{v}_{\pm 2\mathbf{q}}$, drop out, since the velocity is solenoidal in the smectic. Hence, the flow velocity $\mathbf{v}$ will only couple to the transverse part of the perturbation in $\psi$. Hydrodynamic effects do not affect the stability of $\psi$ for longitudinal distortions of the layers.

As we are interested in the transverse stability through modulations of the phase, we impose a perturbation in the plane orthogonal to the layering normal, say $\mathbf{Q} = Q \hat{x}$. By substituting $\mathbf{f}_\mathbf{Q}$ into Eq. (\ref{eq:vel-stb}), we find a velocity in the longitudinal direction $\hat{z}$ that only depends on the Fourier transform on the $z$ component of the forcing term, $(f_z)_{\pm\mathbf{Q}}$. Finally, by substituting this expression for the velocity into Eq. (\ref{eq:psi-stb}) and gathering terms associated with modes exp($\pm i(\mathbf{q}\pm\mathbf{Q}\cdot\mathbf{x})$), we obtain the amplitude equations that govern the evolution of $A_1$ and $A_2$,
\begin{eqnarray*}
\partial_t A_1 &=& (6\beta A_0^2 - 30\gamma A_0^4 + \epsilon - l_1^2 - H(l_1^2-l_0^2))A_1 
\\[2mm]
&&\hspace{25mm}+ (H(l_2^2-l_0^2)+3\beta A_0^2 - 20\gamma A_0^4) A_2^* \;,
\\[3mm]
\partial_t A_2 &=& (6\beta A_0^2 - 30\gamma A_0^4 + \epsilon - l_2^2 - H(l_2^2-l_0^2))A_2 
\\[2mm]
&&\hspace{25mm}+ (H(l_1^2-l_0^2)+3\beta A_0^2 - 20\gamma A_0^4) A_1^* \;.
\end{eqnarray*}
where $H$ is a hydrodynamic coupling coefficient obtained from Eq. (\ref{eq:vel-stb}), and is given by
\begin{eqnarray*}
H &=& \frac{1}{\eta|\mathbf{Q}|^2}\bigg(|\mathbf{q}|^2-\frac{(\mathbf{q}\cdot\mathbf{Q})^2}{|\mathbf{Q}|^2}\bigg)A_0^2 \; .
\end{eqnarray*}

As previously argued, there is no hydrodynamic coupling for longitudinal perturbations, so that $H = 0$ when $\mathbf{q}$ and $\mathbf{Q}$ are parallel. However, for the case of transverse perturbations, the coupling coefficient can have an significant role in the stability. To derive the dispersion relation we need a solution for $A_0^2$ (note that $A_0$ is constant and real), found by gathering terms in the base wavenumber $q_0$,
\begin{eqnarray*}
0 &=& -\epsilon A_0 - \alpha\, l_0^2 A_0 + 3\beta A_0^3 - 10 \gamma A_0^5
\end{eqnarray*}
which gives us
\begin{eqnarray}
A_0^2 &=& \frac{3\beta + \sqrt{9\beta^2-40\epsilon\gamma}}{20\gamma}\;.
\label{eq:a0}
\end{eqnarray}

For transverse modulations of the phase, considering a base frequency $\mathbf{q} = q_0\hat{z}$ and perturbation $\mathbf{Q} = Q\hat{x}$, we can write the order parameter is $\psi = 2 A_0 \textrm{cos} (q_0z + \phi\, \textrm{sin} (Qx))$, where $\phi$ is the amplitude of the initial perturbation. In the limit of $\phi \ll 1$, this is equivalent to setting $A_1 = A_0 \phi/2$ and $A_2 = A_0 \phi/2$ in Eq. (\ref{eq:psi-dist}). By substituting $A_0$, $A_1$, and $A_2$ into the amplitude equation for $A_1$ above, we find
\begin{eqnarray}
\partial_t \phi &=& -\rho_s \bigg[\frac{(3\beta+\sqrt{9\beta^2-40\epsilon\gamma})}{10\eta\gamma}q_0^2Q^2 + Q^4 \bigg]\phi \;,
\label{eq:dispersion}
\end{eqnarray}
The decay rate for the transverse perturbation is given by $\sigma_\perp = \partial_t\phi/\phi$. Therefore, while in the absence of hydrodynamics the growth rate is proportional to $Q^4$, hydrodynamic effects lead to a decay proportional to $Q^{2}$ at low wavenumbers.

\subsection{Code validation}

We compare the numerical solution of the model using the numerical method described in Sec. \ref{sec:nm} to the dispersion of Eq. (\ref{eq:dispersion}). We set the viscosity to be small, $\eta = 0.1$, so that we are able to distinguish the effects from order parameter relaxation and flow. The parameters of the model used are $\beta = 0.4$, $\gamma = 3$, and $\epsilon = \epsilon_c = 0.009$. The base amplitude $A_0$ is computed from Eq. (\ref{eq:a0}), and the perturbation amplitude is $\phi = 0.1$. We use $N = 512^3$ and $\Delta x = 0.7854$ (8 grid nodes per wavelength) and $\Delta t = 5\times 10^{-4}$.

\begin{figure}[ht]
    \centering
    \includegraphics[width=8cm, height=8cm]{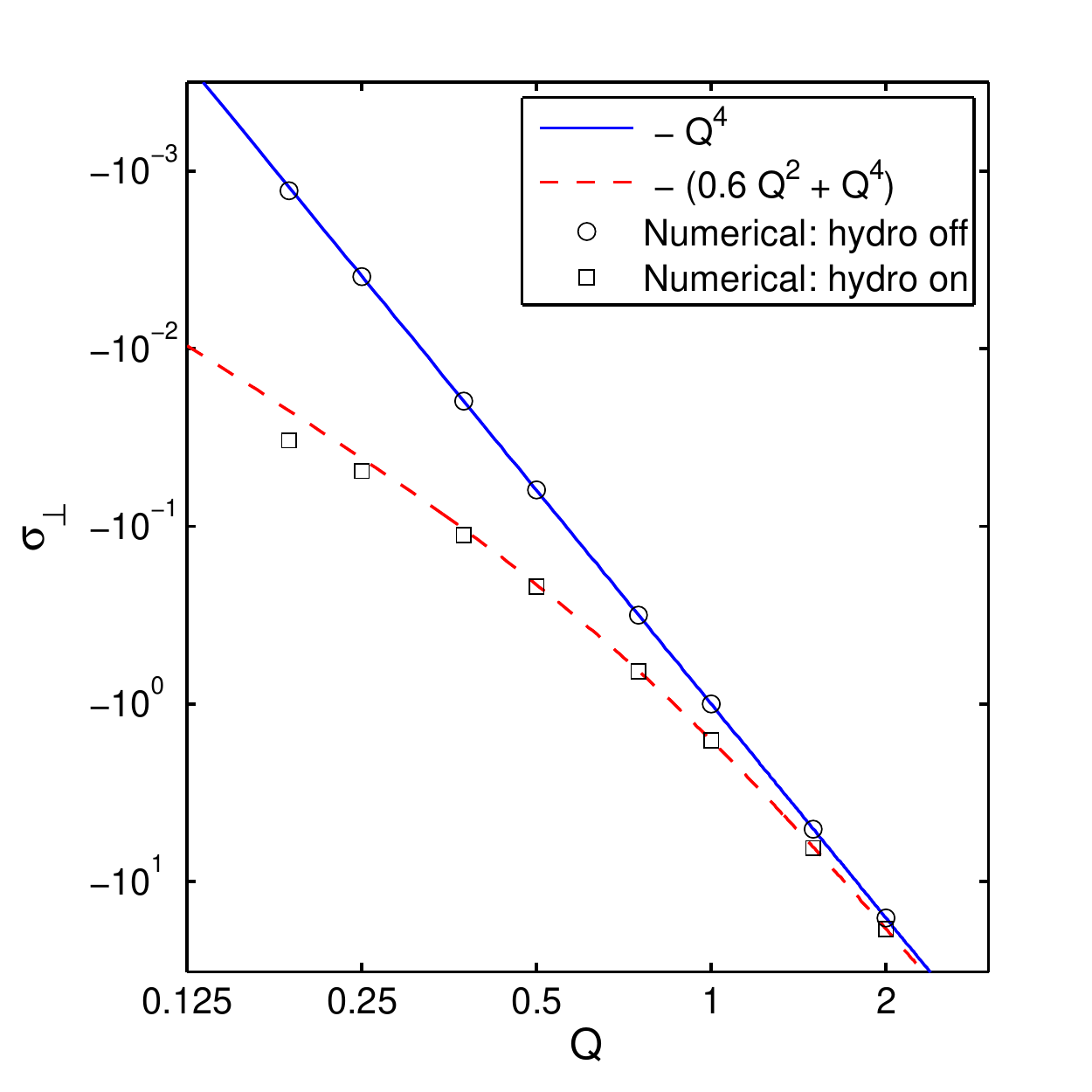}
    \caption{Logarithmic plot for the transverse growth rate $\sigma_\perp$ as a function of the perturbation frequency $Q$, showing how numerical results match the analytic predictions. The solid curve represents the hydrodynamic free case, and the dashed curved the case when hydrodynamics is turned on, with viscosity $\eta = 0.1$. Parameters are $\epsilon = 0.009$, $\beta = 0.4$, $\gamma = 3$ and $q_0 = 1$.
    }
    \label{fig:stability}
\end{figure}

We use an initial condition of the form of Eq. (\ref{eq:psi-dist}) with $A_1 = A_0\phi/2$, $A_2 = -A_0\phi/2$, and set the density of the smectic $\rho_s = 1$. The base and perturbation wavenumbers are $\mathbf{q} = q_0\hat{z}$ and $\mathbf{Q} = Q\hat{x}$. Equations (\ref{eq:quasi-blm}) and (\ref{eq:quasi-psi}) are integrated in time, and the growth rate is computed after a few time steps ($\approx 10$). Since we employ the discrete cosine transform for $\psi$, the growth rate is obtained from the spectrum of the transformed $\psi$, by computing the time derivative of the amplitude associated with the frequency $\mathbf{q}\pm\mathbf{Q}$ and dividing by the same amplitude. The results are shown in Fig. \ref{fig:stability}, where we include, as a reference, the decay rate in the absence of hydrodynamic coupling.

We observe that in both cases numerical results agree very well with the analytic prediction from Eq. (\ref{eq:dispersion}). In the hydrodynamic free case, numerical results for the decay rate follow the $-Q^4$ dependence. When hydrodynamic coupling is included, we also obtain a good agreement between numerical results and the derived dispersion relation for all values of $Q$. The amplitude $0.6$ in the figure follows from substitution of the given model parameters into Eq. (\ref{eq:dispersion}). There are no adjustable parameters in this figure.

\subsection{Energy relaxation and stability of the algorithm}
\label{sec:energy}

\begin{figure}[ht]
	\centering
    \begin{subfigure}[b]{0.48\textwidth}
    \includegraphics[width=\textwidth]{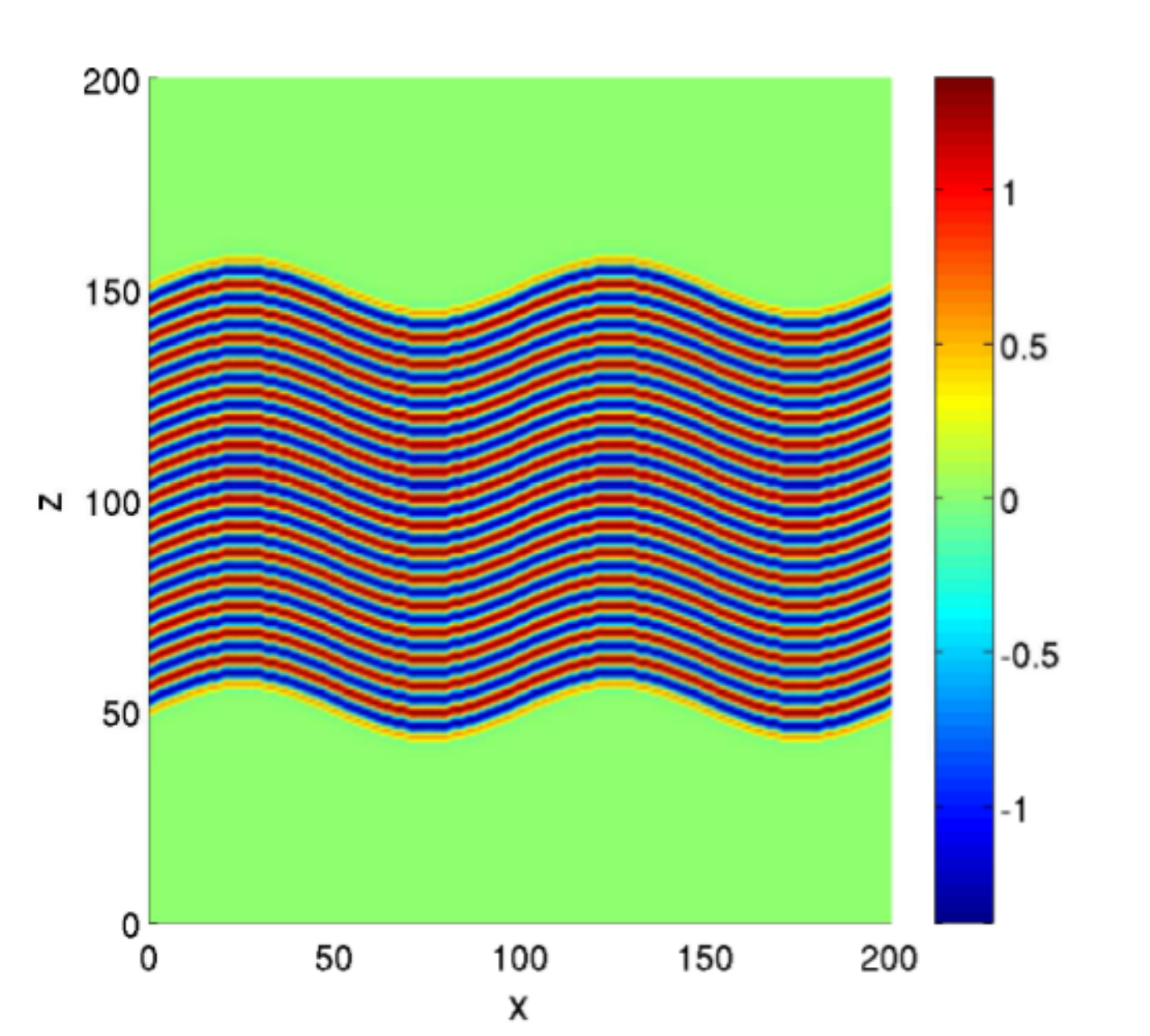}
    \caption{t = 0}
    \label{fig:sma-decay-a}
    \end{subfigure}
    \begin{subfigure}[b]{0.48\textwidth}
    \includegraphics[width=\textwidth]{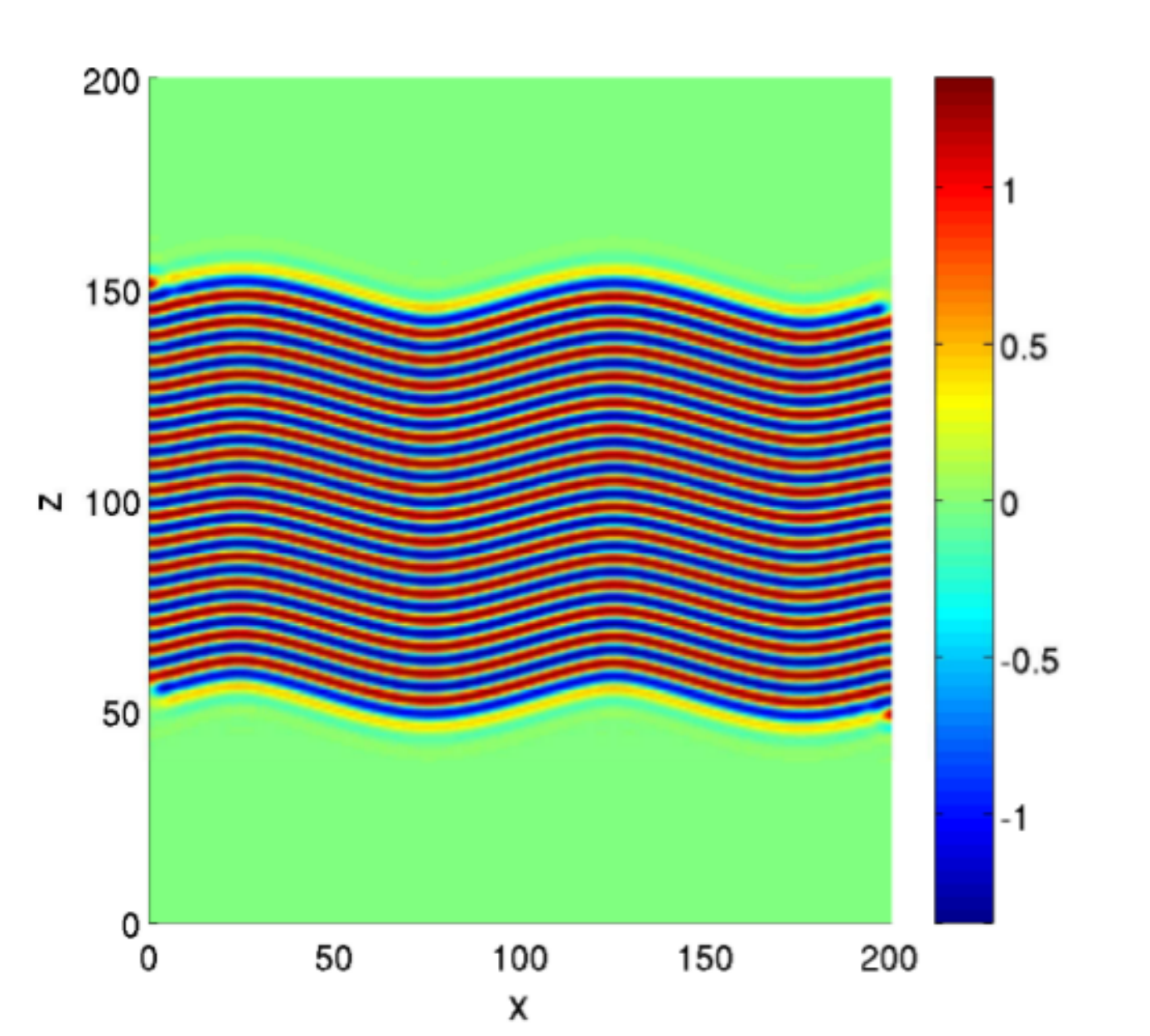}
    \caption{t = 100}
    \label{fig:sma-decay-b}
    \end{subfigure}
    \caption{Two dimensional cross section of an initially perturbed stack of smectic layers in contact with an isotropic fluid relaxing towards a planar configuration while decreasing the total energy of the system.}
	\label{fig:sma-decay}
\end{figure}

The stability of the numerical integration with respect to time step is now investigated by monitoring the decay of the total energy of the system, Eq. (\ref{eq:energy}), with $e$ as defined in Eq. (\ref{eq:energy-density}). The case investigated concerns a slab of distorted smectic planes surrounded by an isotropic fluid at coexistence. We take the smectic layers aligned along the $z$ direction, and perturbed along $\mathbf{Q} = Q\hat{x}$ as in Eq. (\ref{eq:psi-dist}), and as shown in Fig. \ref{fig:sma-decay-a}. The density of the bulk smectic is chosen as $\rho_s \approx 0.67$, and the density of the isotropic fluid $\rho_0 = 0.05$. The parameters in the energy are $q_0 = 1$, $\alpha = 1$, $\beta = 2$, $\gamma = 1$, $\epsilon = \epsilon_c =  0.675$ so that the two phases have approximately the same energy (coexistence). Up to a certain finite value of the perturbation amplitude, we expect the reference planar configuration to be stable, so that the perturbed smectic planes relax as shown in Fig. \ref{fig:sma-decay-b}.

\begin{figure}[ht]
    \centering
    \includegraphics[width=8cm, height=8cm]{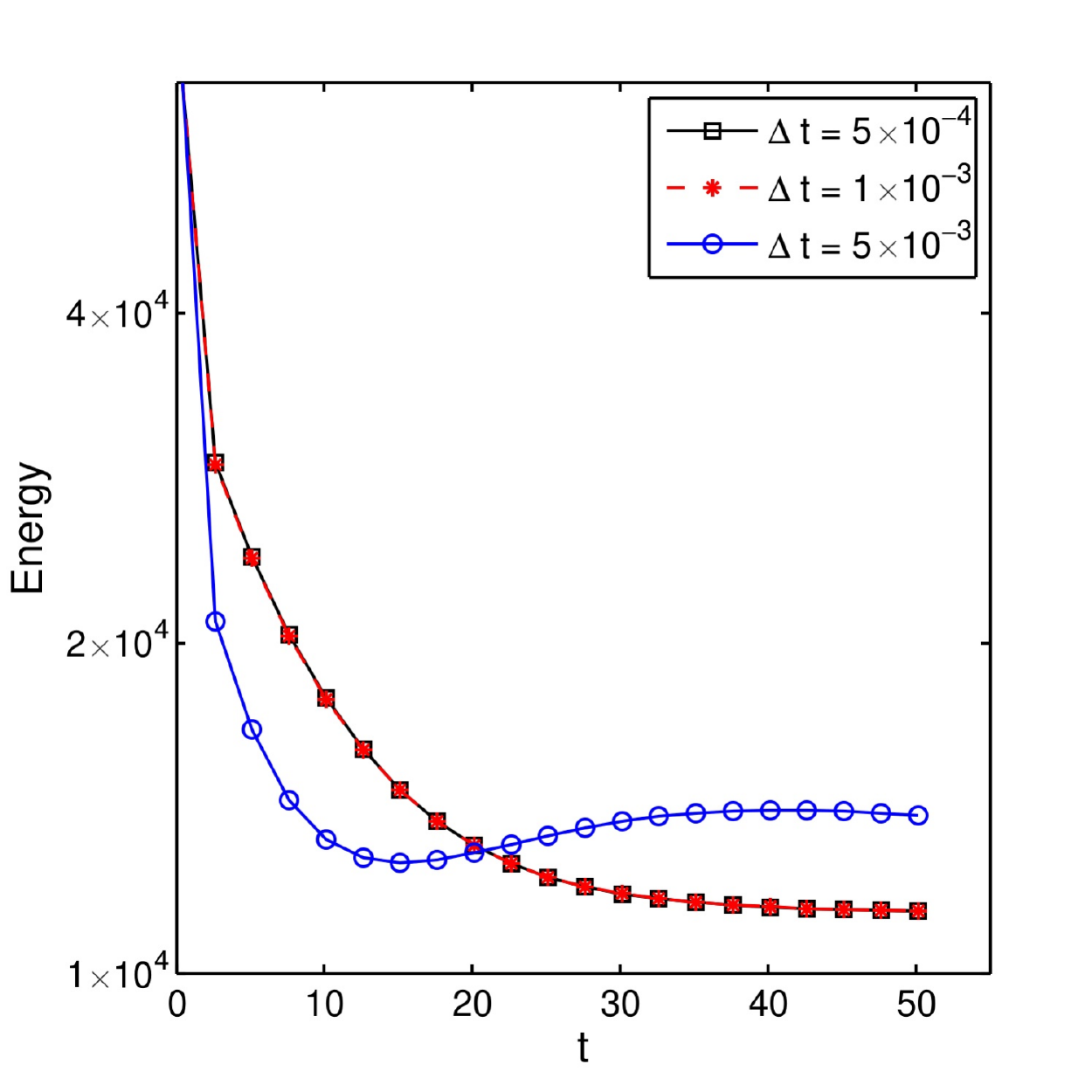}
    \caption{Energy $\int \rho\,e\,dx$ decay in time for different values of the time step $\Delta t$. The curves agree for $\Delta t \leq 1\times 10^{-3}$, while for larger steps the curves deviate.}
    \label{fig:energy}
\end{figure}

We set $\Delta x = 2\pi/8 =  0.7854$ and $N = 256^3$. Time steps are chosen for each of the runs, and we let the system evolve in time, so that the total energy decay can be monitored. Results are shown in Fig. \ref{fig:energy} for three different time steps: $\Delta t = 5 \times 10^{-4}$, $\Delta t = 1 \times 10^{-3}$, and $\Delta t = 5 \times 10^{-3}$. We observe that the curves match for $\Delta t = 5 \times 10^{-4}$ and $\Delta t = 1 \times 10^{-3}$, and both exhibit the expected monotonic decay. We obtain the same curves for smaller values of $\Delta t$. However, the curve $\Delta t = 5 \times 10^{-3}$ diverges from the previous ones, and fails to be monotonic. For $\Delta t \geq 1 \times 10^{-2}$, the numerical scheme becomes unstable and numerical solutions diverge.

These results show that the scheme introduced in Sec. \ref{sec:nm} for dealing with a dynamic equation for the order parameter with phases of varying density does not impose overly severe restrictions on time step. For instance, for a smectic density $\rho_s \approx 1$, the time step is of the same order as in the semi-implicit scheme employed for purely diffusive decay in Ref. \cite{vitral2019role}. While $\Delta t = 1 \times 10^{-3}$ is an appropriate choice for the time step in this case, some factors may require this choice to be altered. For example, increasing the resolution to have more points representing the base wavelength requires $\Delta t$ to be decreased. Another factor is associated with the balance of mass from Eq. (\ref{eq:quasi-divv}): by increasing the difference ratio between the smectic and disordered phase densities, numerical instabilities may arise from the way $\nabla\dot{\bf v}$ is computed from the material time derivative of the amplitude over the density. Hence, the appropriate choice of $\Delta t$ and $\Delta x$ must be done on a case to case basis.

Finally, we conclude this section by mentioning that we have checked that the numerical method conserves mass at coexistence of phases. For the same initial condition (i.e., transversely perturbed smectic layers), parameters as above, and $\Delta t = 1\times 10^{-3}$, we have followed how the mass fraction $m/m_0$ changes in time, where $m_0$ is the initial mass, and $m = \int \rho dV$ is computed after every time step. While there is a slight decay of mass (approx. $2\%$) at the start due to relaxation of the imposed initial condition, mass gradually returns towards its initial value. For long times ($t > 60$), mass reaches a constant value, at a mass fraction $m/m_0$ of $99.8\%$.


\section{Flow structure in smectic-isotropic fluid configurations}
\label{sec:flow}

The balance of linear momentum can be written in terms of a body force $\mathbf{f} = \mu\nabla\psi$, as seen in Eq. (\ref{eq:blm-stb}). The force $\mathbf{f}$ is zero either for planar smectic layers, or at coexistence with the isotropic fluid across a planar interface. For curved layers parallel to a curved interface, the chemical potential with respect to planarity $\delta\mu$ becomes a function of the curvatures of the surfaces of constant $\psi$, and is given by an extension of the Gibbs-Thomson equation \cite{vitral2019role}
\begin{eqnarray}
    \delta\mu\Delta A &=& 2H\sigma_h + (4H^2-2G)\sigma_b - 2H(3G-4H^2)\sigma_t \;.
    \label{eq:gibbs-thomson}
\end{eqnarray}
where $\Delta A$ is the difference in amplitude between the smectic and the isotropic phases, $\sigma_h$ is the surface tension, $\sigma_b$ the interface bending coefficient, and $\sigma_t$ the interface torsion coefficient. These three coefficients can be obtained analytically from the model parameters and the solution for the amplitude $A$ corresponding to a stationary amplitude across a planar interface \cite{vitral2019role}. The mean curvature $H = (c_1+c_2)/2$ is the average of the principal curvatures $c_1$ and $c_2$ at a surface point, while the Gaussian curvature $G = c_1 c_2$ is the product of these curvatures. The factor $\nabla \psi$ in $\mathbf{f}$ ensures that the force is normal to the smectic layers. At the interface, a positive normal points outwards away from smectic, and the sign of $H$ is such that it is positive for a sphere. At the interface, given that the amplitude $A$ goes from its finite value in the bulk smectic to zero in the isotropic phase, to lowest order in curvature the force $\mathbf{f}$ is directed towards (resp. away) from the nearest center of curvature when $\delta \mu > 0$ (resp. $\delta \mu < 0$), so that at an interface it points towards the smectic phase in regions of positive $H$.

\begin{figure}[ht]
	\centering
    \begin{subfigure}[b]{0.48\textwidth}
    \includegraphics[width=\textwidth]{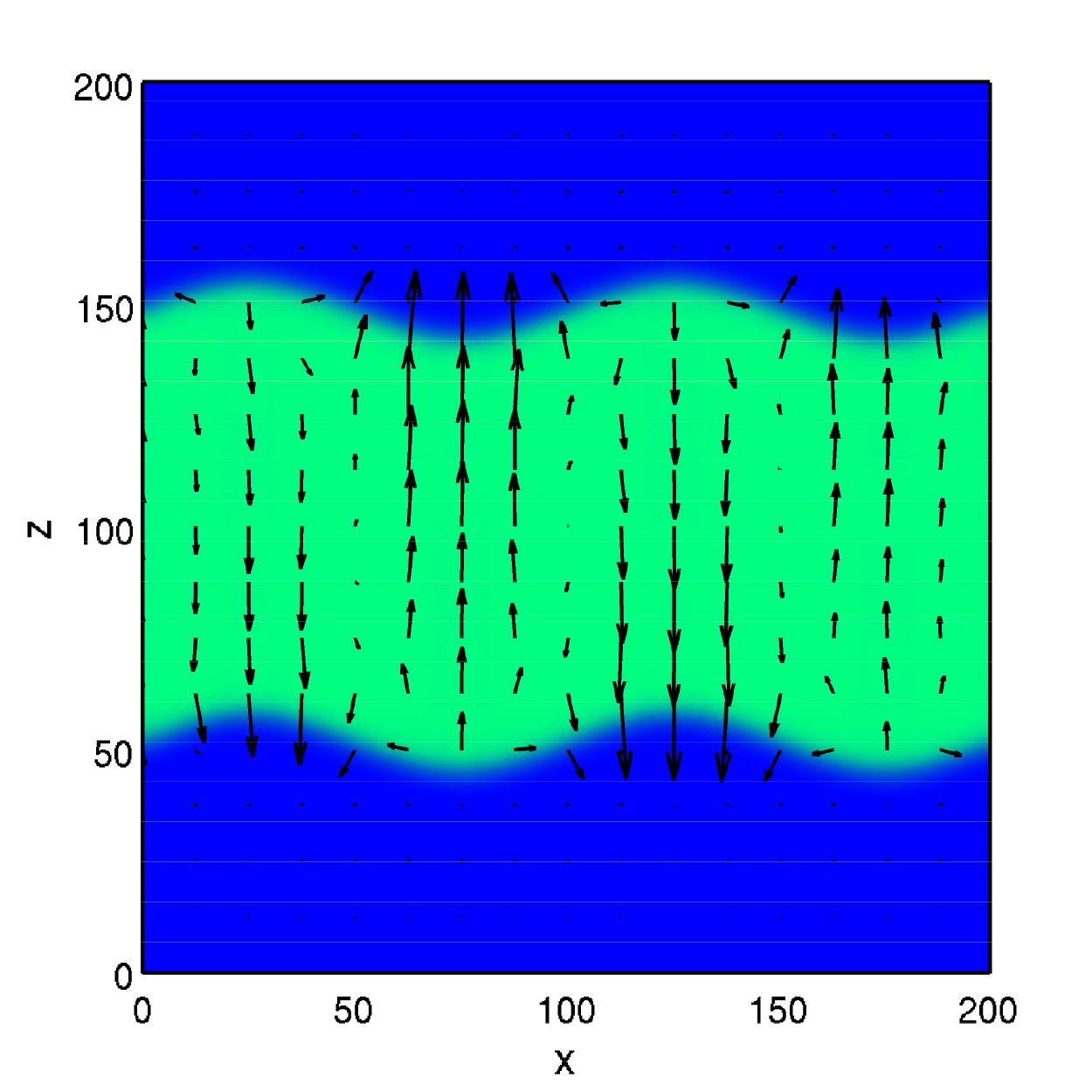}
    \caption{Perturbed smectic}
    \end{subfigure}
    \begin{subfigure}[b]{0.48\textwidth}
    \includegraphics[width=\textwidth]{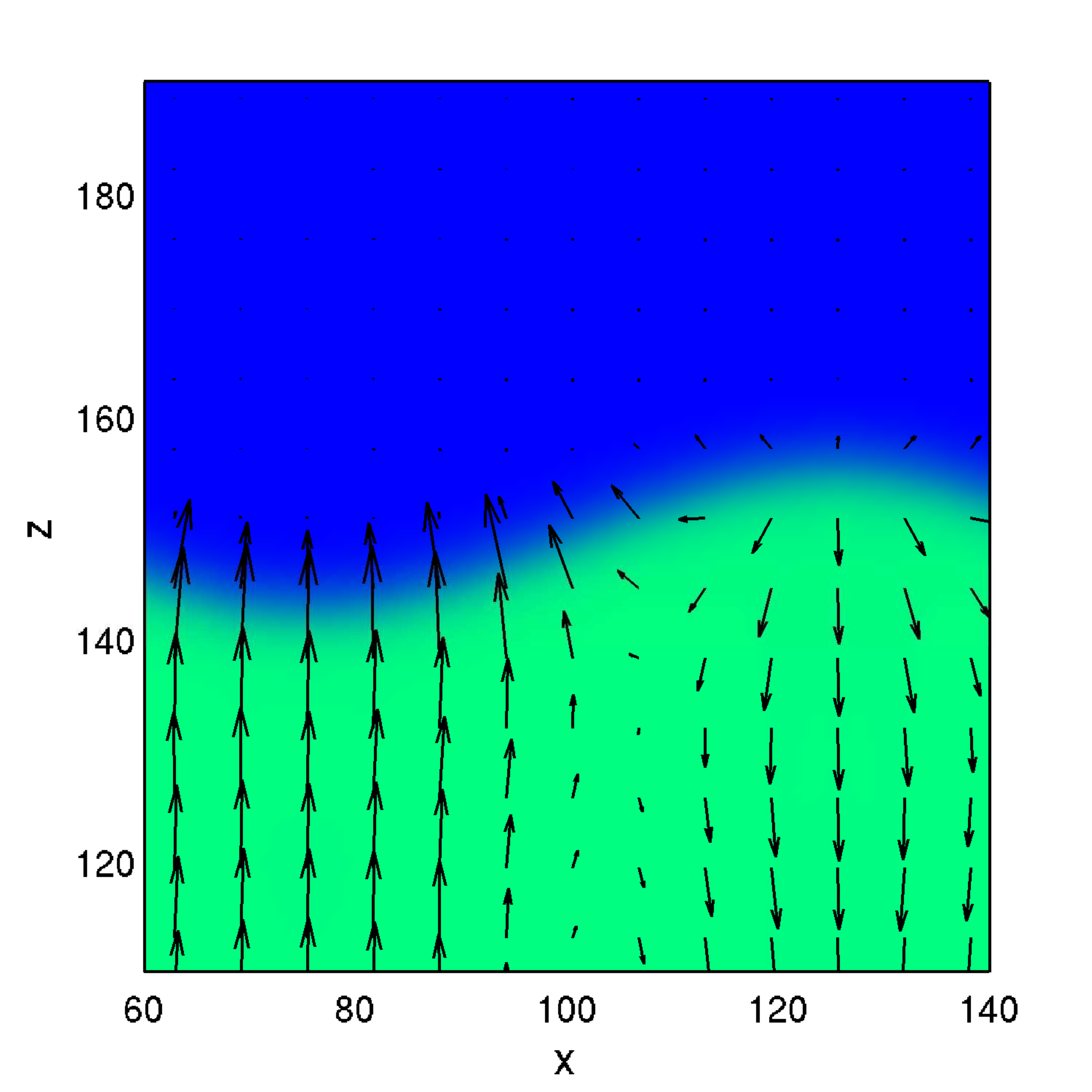}
    \caption{Enlarged}
    \end{subfigure}
    \caption{Mass flow $\rho \mathbf{v}$ and density field $\rho$ at time for a transversely perturbed smectic at $t = 2$, with a density ratio of approximately 10:1. The right image is a magnification of the left one, showing the flow structure near the interface. We use $N = 256^3$, $\Delta t = 5\times 10^{-4}$, and parameters $\kappa = 0.5$, $\rho_0 = 0.05$, $q_0=1$, $\eta=10$, $\epsilon = 0.675$, $\alpha=1$, $\beta=2$ and $\gamma=1$.}
	\label{fig:quiver}
\end{figure}

With these considerations in mind, we present results on the structure of the flow for two different configurations: a transversely modulated smectic layer in contact with the isotropic fluid as in Sec. \ref{sec:energy}, and a smectic domain in the form of a focal conic. We consider the following values of the model parameters: $\kappa = 0.5$, $\rho_0 = 0.05$ (density ratio above 10:1), $q_0 = 1$, $\alpha = 1$, $\beta = 2$, $\gamma = 1$, $\epsilon = \epsilon_c = 0.675$, and  viscosity is $\eta = 10$.  For this value of the viscosity, the non-solenoidal velocity has a strong contribution to the resulting interfacial flows. Figure \ref{fig:quiver} shows the transient mass flux, $\mathbf{v}_m = \rho \mathbf{v}$, alongside the density field (green for high density, smectic, and blue for low density fluid), for time $t = 2$. On the interface, we observe that mass flows outward from smectic regions of negative mean curvature (growth), while in regions of positive mean curvature mass flows inwards towards the smectic phase and also towards regions of negative mean curvature. This is in agreement with our discussion about the direction of the force $\mathbf{f}$ as a function of curvatures.

One of the main motivations of our study is to udenrstand the stability and evolution of focal conic domains in smectics films. A focal conic exhibits a macroscopic singularity at its center, hence the phase field approach is well suited to study this configuration. Three important effects relevant to focal conics are captured by our model. First, non classical stresses are present at the interface between the conic and the isotropic fluid that depend on both mean and Gaussian curvatures. Second, given a density contrast between the smectic and surrounding fluid, a non-solenoidal velocity field at the interface can introduce significant changes to mass transport and therefore to flow structure and stability. Finally, as shown in \cite{vitral2019role} (but not in the example below), instability of a smectic-fluid interface can result in exposed smectic layers at the interface. Their local evolution is governed by Willmore type flows instead of capillarity driven flows. 

\begin{figure}[ht]
	\centering
    \begin{subfigure}[b]{0.48\textwidth}
    \includegraphics[width=\textwidth]{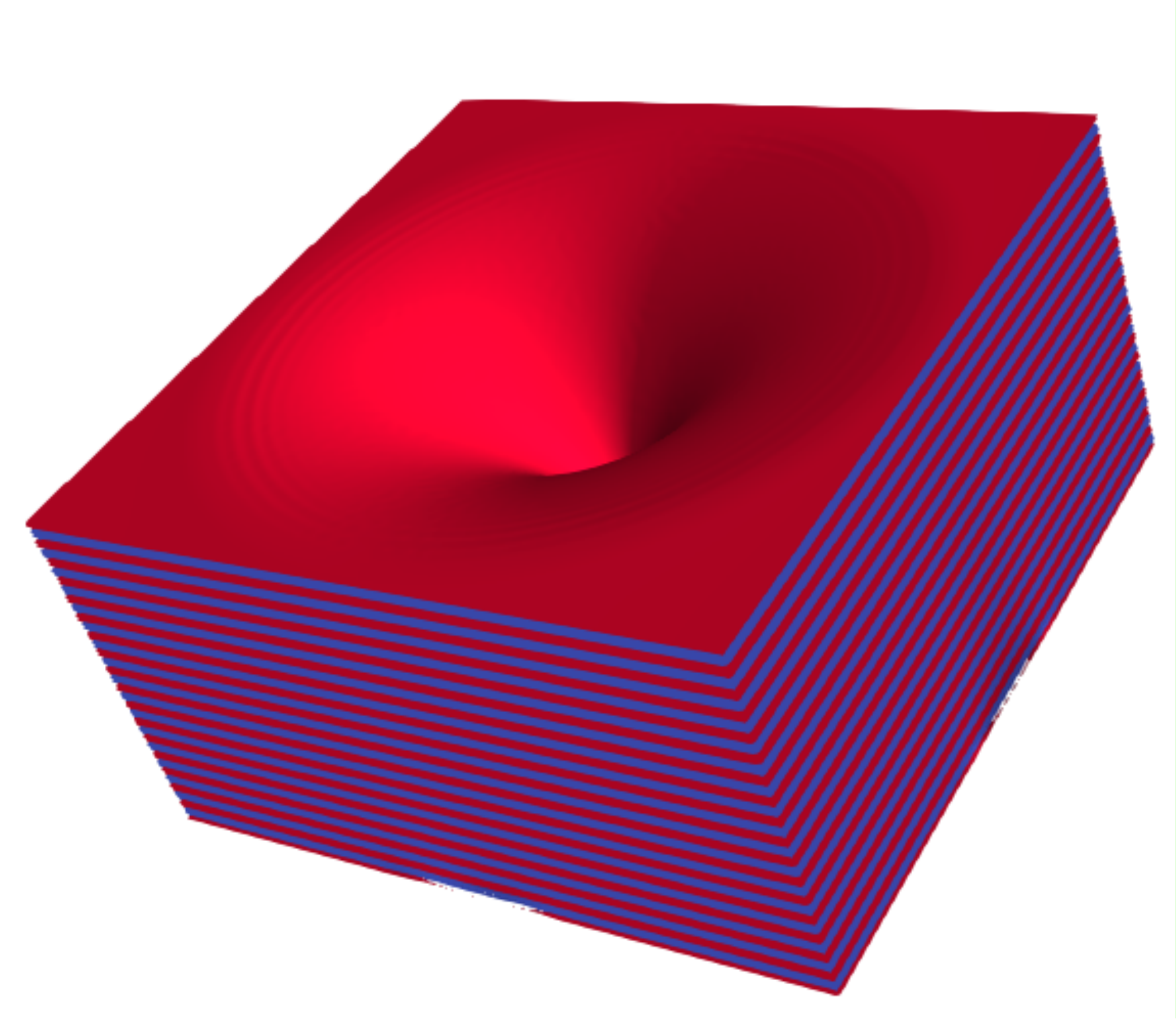}
    \caption{Focal conic}
    \end{subfigure}
    \begin{subfigure}[b]{0.48\textwidth}
    \includegraphics[width=\textwidth]{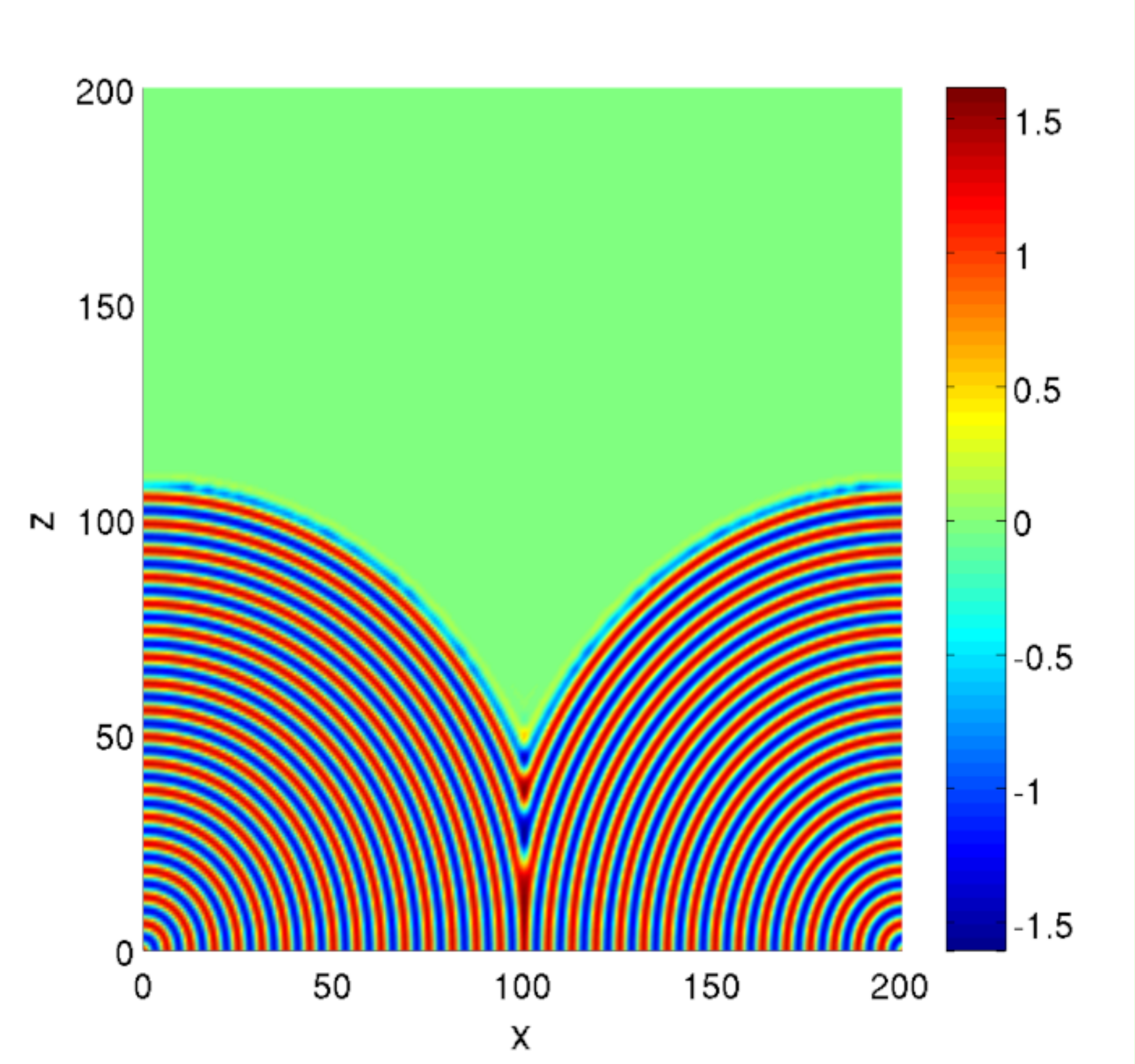}
    \caption{Cross-section}
    \end{subfigure}
    \caption{Smectic layers bent in a focal conic configuration. (a) The color code represents the order parameter  $\psi$ ranging between $\psi_{max}$ (red) and $\psi_{min}$ (blue). (b) Middle cross-section displaying the values the order parameter field.}
	\label{fig:fc}
\end{figure}

\begin{figure}[ht]
	\centering
    \begin{subfigure}[b]{0.48\textwidth}
    \includegraphics[width=\textwidth]{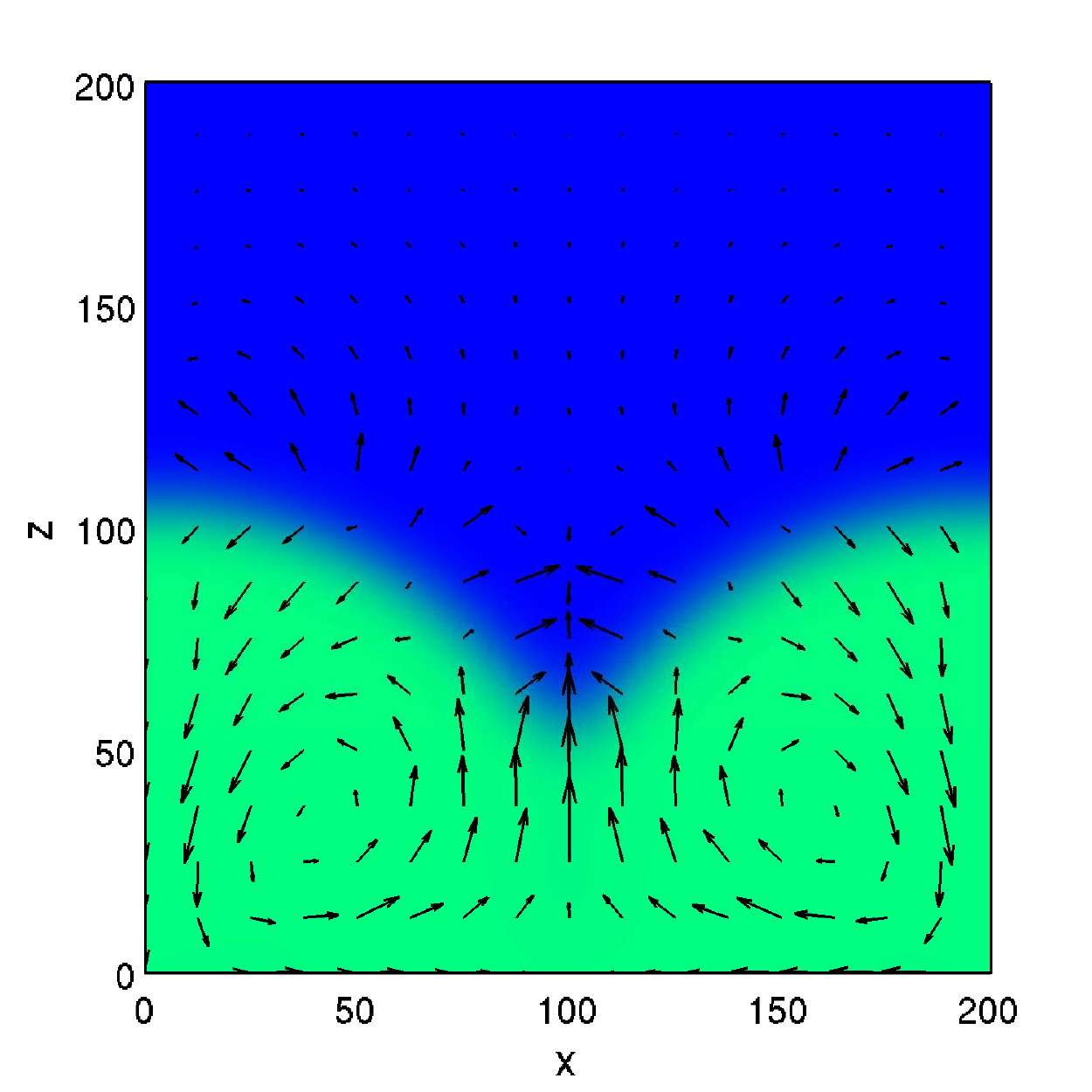}
    \caption{Density ratio 100:1}
    \end{subfigure}
    \begin{subfigure}[b]{0.48\textwidth}
    \includegraphics[width=\textwidth]{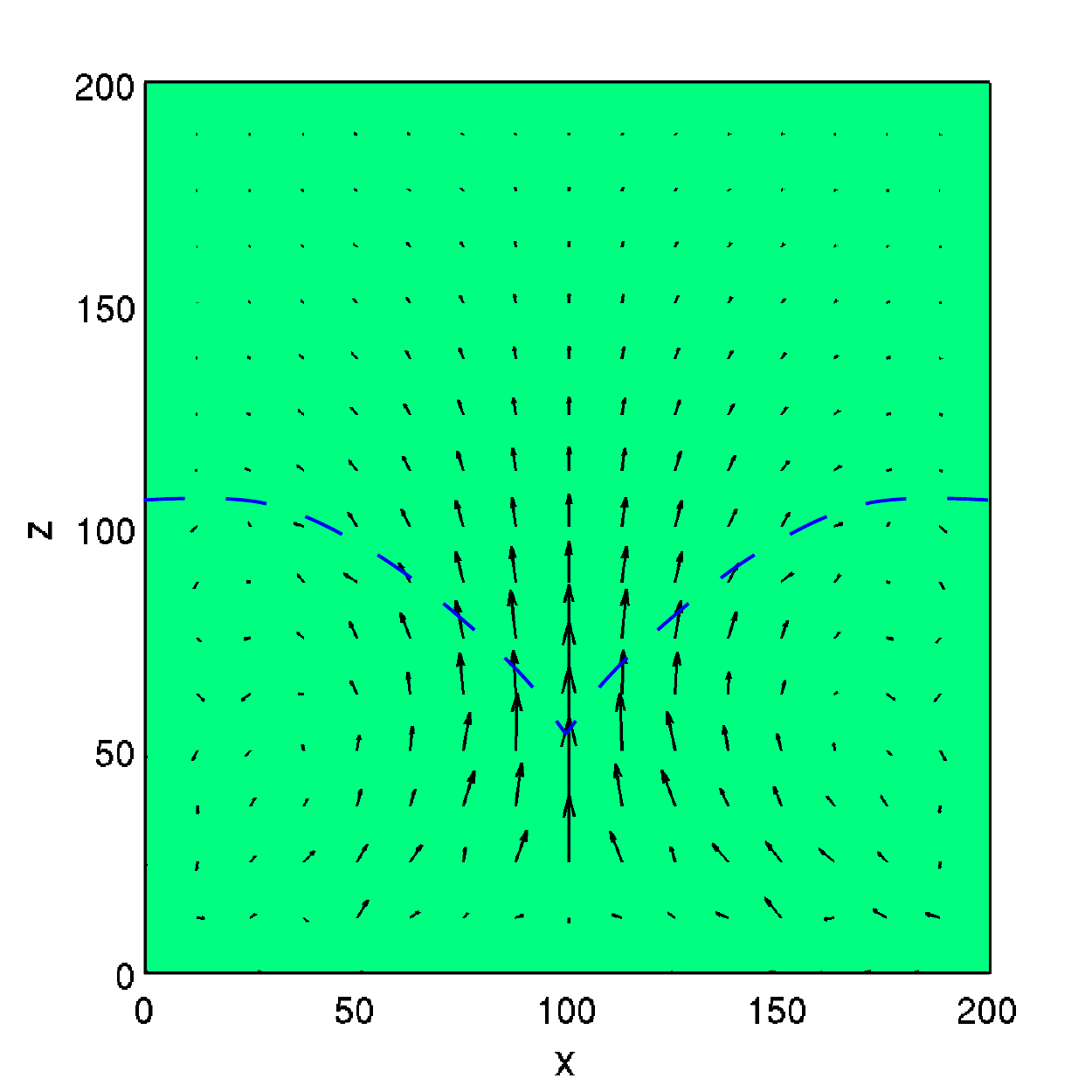}
    \caption{Uniform density}
    \end{subfigure}
    \caption{Comparison between the transient fluid flow $\mathbf{v}$ on smectic-isotropic fluid system with approximately 100:1 density contrast ($\kappa = 0.5$, $\rho_0 = 0.005$) and on a smectic-isotropic system of homogeneous density, where the dashed lines mark the location of the interface. The density is represented by the background color: green for high density and blue for low density. We use $N = 256^3$, $\Delta t = 5\times 10^{-4}$, and parameters $q_0=1$, $\eta=1$, $\epsilon = 0.675$, $\alpha=1$, $\beta=2$ and $\gamma=1$.}
	\label{fig:fcflow}
\end{figure}

We show in Fig. \ref{fig:fc} an initial configuration comprising a focal conic domain in three dimensions surrounded by an isotropic fluid of different density. 
We use the same model parameters as in the previous case except that $\rho_0 = 0.005$, so that we have a density ratio greater than 100:1 between the smectic and the isotropic fluid. We also set $\eta = 1$. Figure \ref{fig:fcflow} shows the transient velocity field at time $t = 5$ at the center cross section, with average velocity $\mathcal{O}(10^{-1})$, alongside the density in the background (green for high density, blue for low). The flow pushes the smectic outwards near the conic center, which is a region of negative mean curvature, while away from the core it pushes inwards, as it is a region of positive mean curvature. This is in agreement with the dependency of the force $\mathbf{f}$ with respect to Eq. (\ref{eq:gibbs-thomson}) to lowest order in curvatures. Hence within the incompressible smectic we observe a recirculating toroidal flow. Given the large density contrast between the phases, the flow velocity exhibits a large variation for most of the interface (away from the center). For comparison, we also show results for the same focal conic for a smectic-isotropic interface when the density is uniform, and for the same dimensionless time $t = 5$. It is clear that the flow is continuous on the smectic-isotropic interface since the velocity is solenoidal. While we also observe the flow moving outward on regions of negative mean curvature, and inward on regions of positive mean curvature, it does so continuously through the smectic-isotropic transition creating advection rolls that span the two phases, as expected for this fully incompressible case.

\begin{figure}[ht]
    \centering
    \includegraphics[width=0.55\textwidth]{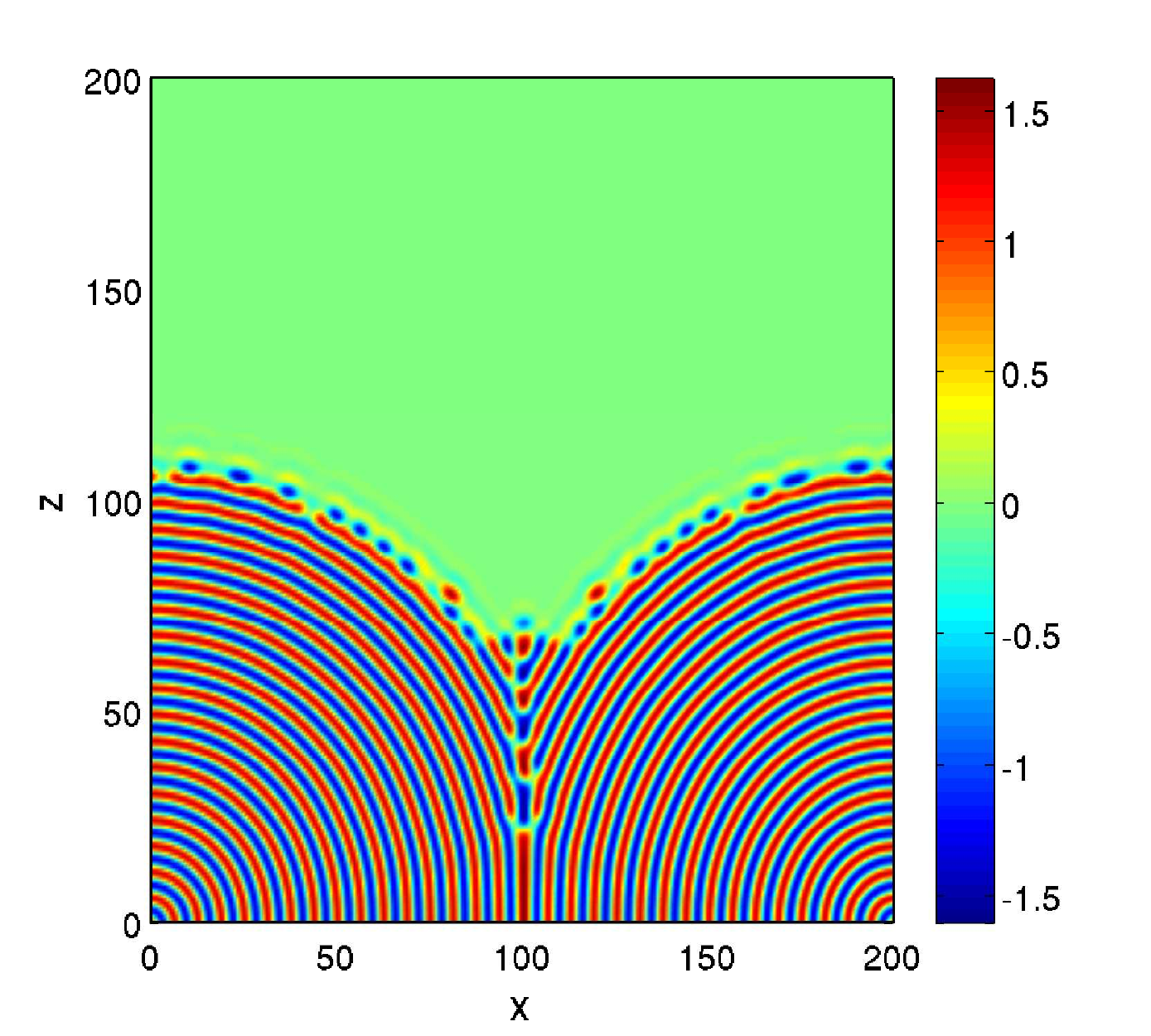}
    \caption{Focal conic configuration at time $t = 310$, using the same parameters as in Fig. \ref{fig:fc}, for a density ratio of approximately 100:1. As in the purely diffusional dynamics, rings are formed at the top of the focal conic. Additional breakage of layers close to the interface and regions of self-intersection are due to the order parameter advection. The dynamics are very slow at this stage, but it is not a steady state configuration.}
    \label{fig:fcrho100to1t310}
\end{figure}

When the focal conic configuration evolves for long times, we observe that the flow considerably slows down in the smectic, away from the central vertical axis where layers self-intersect.  In the case of purely diffusional dynamics (relaxational motion of $\psi$), the focal conic slowly evolves toward a steady state configuration \cite{vitral2019role}, creating rings at the surface and filling the singularity with smectic.  In contrast, in the presence of flow, we observe flow-induced corrections to the resulting morphologies. Fig. \ref{fig:fcrho100to1t310} shows the morphology obtained by letting the focal conic from Fig. \ref{fig:fcflow} evolve up to time $t = 310$, keeping the approximate 100:1 density contrast. In addition to the creation of rings at the smectic-isotropic interface (small blue disks in the cross-section), we find that flow induces the breaking of layers right below the rings, which we did not observe in the absence of flow.  Also, we see that different layers break in the vicinity of their self-intersection point, which again does not happen for purely diffusional dynamics. This observation is connected to the high curvatures and deviations from the layering wavenumber in regions where layers self-intersect, generating stress and inducing flows that leads to ruptures in these internal layers. While the dynamics at this stage becomes very slow, fluid flow is still relevant, with average velocity $\mathcal{O}(10^{-2})$, and advection of the order parameter may induce more rupture of layers for even longer times.

\begin{figure}[ht]
	\centering
    \begin{subfigure}[b]{0.48\textwidth}
    \includegraphics[width=\textwidth]{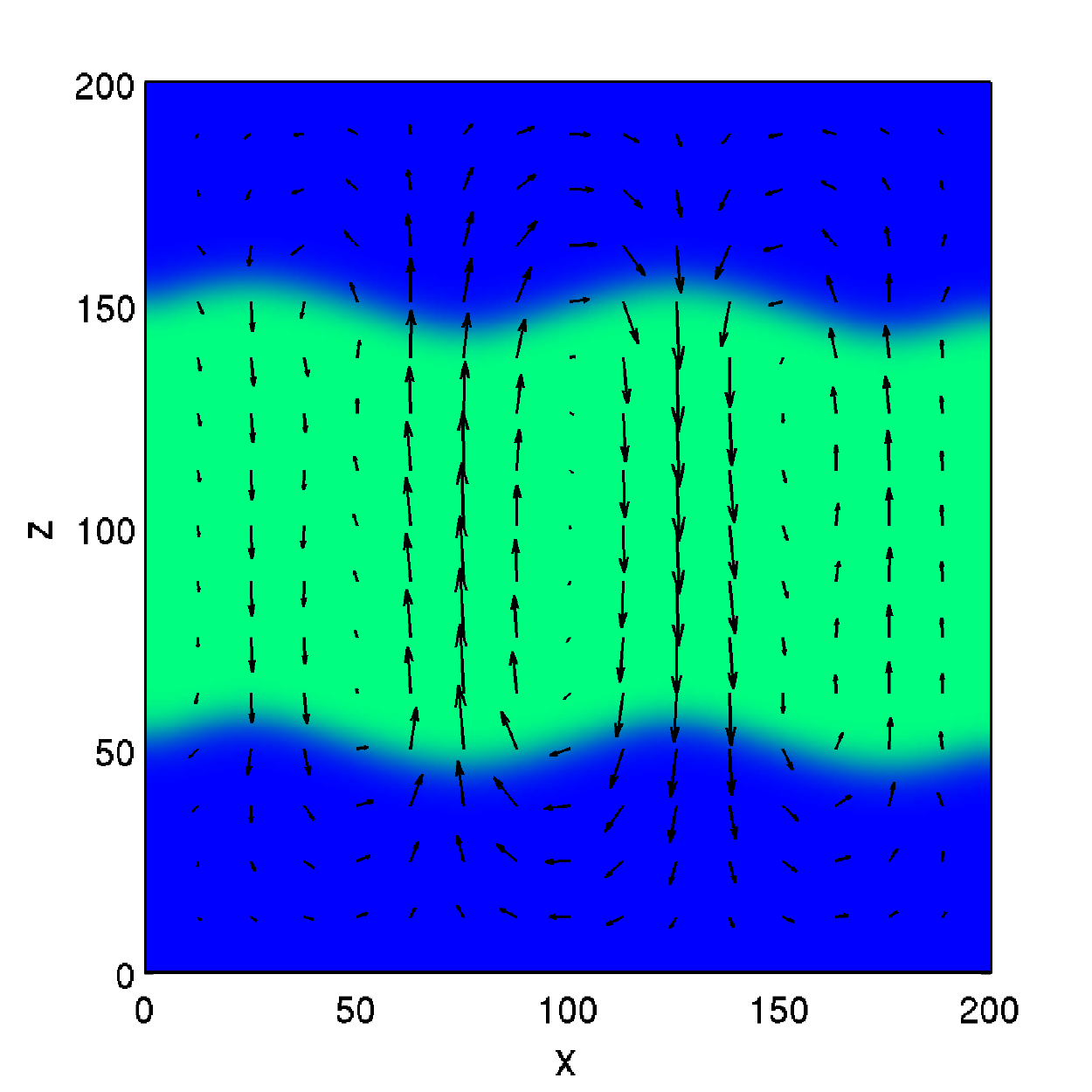}
    \caption{Perturbed smectic A, $(\rho_s-\rho_0)/\rho_s = 0.01$}
    \end{subfigure}
    \begin{subfigure}[b]{0.48\textwidth}
    \includegraphics[width=\textwidth]{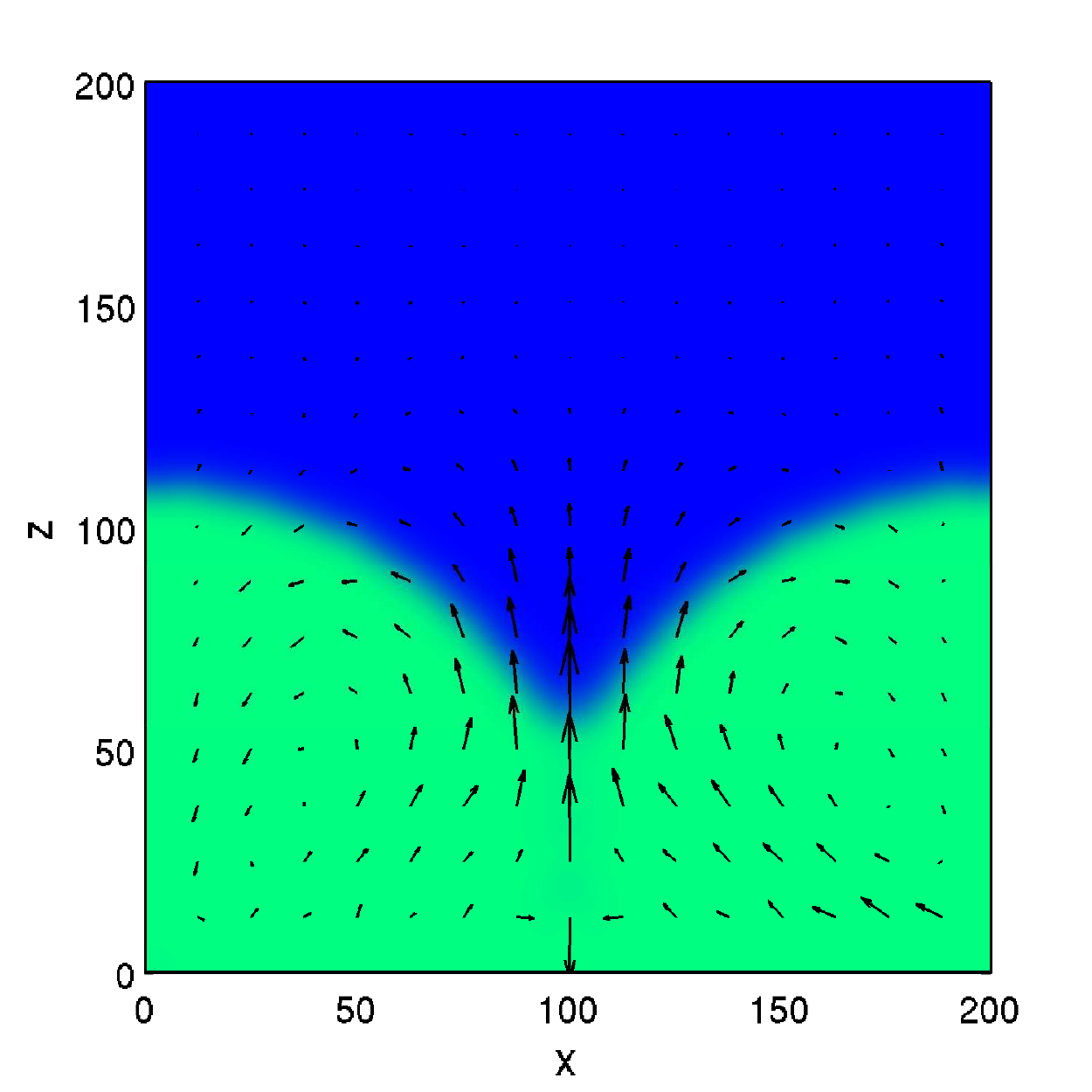}
    \caption{Focal conic, $(\rho_s-\rho_0)/\rho_s = 0.01$}
    \end{subfigure}
    \caption{Fluid flow on a smectic-isotropic fluid system with a density jump of 1\%, a value in the empirical range of transitions in thermotropic low molecular weight materials. The density is represented by the background color: green for high density and blue for low density. We use $\rho_0 = 0.99$, $\rho_s = 1$, and the remaining parameters are the same as in Fig. \ref{fig:fc}.
    }
	\label{fig:thermotropic}
\end{figure}

An application for this model is the case where the isotropic phase represents a fluid that the smectic transitions into at sufficiently high temperatures. In thermotropic low molecular weight materials, the density ratio for such isotropic - smectic A transitions is very low, ranging between 0.5\% and 2\% \cite{prabhu2004phase}. Therefore, using the same parameters as in Fig. \ref{fig:fcflow}, we also studied a system where a smectic A phase with $\rho_s = 1$ of a thermotropic material coexists with its isotropic phase with $\rho_0 = 0.99$, so that we have a density jump of 1\% between them. In Fig. \ref{fig:thermotropic} we show both the perturbed smectic A (at $t = 2$) configuration used in Fig. \ref{fig:quiver} and the focal conic configuration (at $t = 5$) used in Fig. \ref{fig:fcflow} for the case of a 1\% density jump. Due to the small density ratio between phases, the flow is very similar to the fully incompressible case. For the perturbed smectic, we observe advection rolls generated by the flow moving outward on regions of negative mean curvature, and inward on regions of positive mean curvature. The magnitude of the mass flux does not change in the transition between the phases, in contrast to Fig. \ref{fig:quiver}, where it decays in the isotropic phase. For the focal conic, the flow can be described similarly to the uniform density case from Fig. \ref{fig:fcflow}. However, for 1\% density jump we observe a stronger tangential flow at the interface, and advection rolls start bending closer to the smectic. As mentioned in Sec. \ref{sec:nm}, gradients of the density do not pose a problem to numerical stability for such low density ratios.

\section{Conclusions}

We have derived a compressible phase field model for a two phase smectic-isotropic fluid of varying density by introducing an energy density functional of the smectic order parameter and its gradients. Reversible and irreversible currents are derived from the second law of thermodynamics, leading to the governing dynamical equations. We have specialized our analysis to the case in which the bulk fluids in coexistence are incompressible, but compressibility effects are allowed near the two phase interface. In order to accomplish this, we have introduced a constitutive relation for the density which depends only on the amplitude of the smectic order parameter. Therefore the velocity field is non solenoidal only in the interfacial region.

A semi-implicit numerical method was developed to integrate the governing equations which is based on an earlier scheme for phase-field models with varying mobilities. The algorithm has been implemented in a parallel code so that we can examine relatively large three dimensional configurations. We have also conducted a stability analysis of weakly perturbed smectic planes, and derived the corresponding dispersion relation, Eq. (\ref{eq:dispersion}), for transverse modes. At long wavelengths, hydrodynamic effects dominate the dispersion relation, with a $Q^{2}$ wavenumber dependence, instead of the $Q^{4}$ expected for diffusive decay. We have validated our code against this dispersion relation.

We have presented numerical results concerning fluid flows for a smectic film surrounded by an isotropic fluid of different density. When the initial configuration comprises a set of smectic layers that are weakly perturbed along the transverse direction, we observe that in regions of negative mean curvature at the interface the flow is outward away from the smectic, while those of positive mean curvature push the flow inward toward the smectic, as expected from the dependence of the surface stress on curvature. In a focal conic configuration, flow in the bulk smectic is, for the parameters considered, a convective roll as expected in an incompressible fluid. In both configurations, there is a large variation in velocity across the interface due to the density variation associated with the local gradient of order parameter. We are currently investigating the quantitative effects of higher order curvature terms in the interfacial stress on fluid flow near the conic center, its stability, and its nonlinear evolution. Furthermore, hydrodynamic flows are expected to introduce long-range interactions between focal conics, a subject of considerable interest in applications of arrays of focal conics in smectic films.


\section*{Acknowledgments}

This research has been supported by the Minnesota Supercomputing Institute, and by the Extreme Science and Engineering Discovery Environment (XSEDE) \cite{towns2014xsede}, which is supported by the National Science Foundation under grant number ACI 1548562. EV thanks the support from the Doctoral Dissertation Fellowship and from the Aerospace Engineering and Mechanics department, University of Minnesota. The research of JV is supported by the National Science Foundation under Grant No. DMR-1838977.


\appendix
\section{Derivation of the governing equations in the displacement configuration variable \textit{u}}
\label{sec:ap}

In this section, we derive the governing equations for an incompressible smectic based on the configuration variable $u$, which accounts for layer displacements. As usual, we assume that $e = e(s,\nabla u)$, so that the energy only depends on gradients of $u$ and is invariant under simple translations of the structure. Since this is an energy based on gradients of the layer displacement, and does not describe a two-phase interface problem, we define it as a volumetric energy density. Similarly as before, the balance of internal energy is given by
\begin{eqnarray*}
    \dot{e} &=& \mathbf{T}:\nabla \mathbf{v} + \nabla \cdot (\mathbf{t}\,\dot{u}) + r \; .
\end{eqnarray*}

Following the same steps from Sec. \ref{sec:comp}, we derive the local balance of entropy
\begin{eqnarray*}
    \nonumber
    \theta \dot{s} &=& 
    \bigg( \mathbf{T} +\nabla u\otimes
    \frac{\partial e}{\partial \nabla u}\bigg) : \nabla \mathbf{v}
    +\bigg( \mathbf{t} -\frac{\partial e}{\partial \nabla u}
    \bigg) \cdot \nabla \dot{u} + \nabla\cdot\mathbf{t}\,\dot{u} + r \;.
\end{eqnarray*}

Using the Coleman-Noll procedure, we obtain the reversible currents by setting the entropy production rate to zero, so that
\begin{eqnarray*}
\mathbf{T}^R &=& -\nabla u \otimes \frac{\partial e}{\partial \nabla u} \;,
\\[2mm]
\mathbf{t}^R &=& \frac{\partial e}{\partial \nabla u} \;.
\end{eqnarray*}
The expression obtained for the reversible part of the generalized force $\mathbf{t}$ is the thermodynamic conjugate to $\nabla u$. This derivative plays the role of a molecular field in smectics, which is commonly labeled as $\mathbf{h}$, and whose divergence is a thermodynamic force (in nematics, the molecular field is conjugated to the director $\mathbf{n}$, and at equilibrium $\mathbf{n}$ should be at each point parallel to $\mathbf{h}$ \cite{degennes1995physics}). 

Since $u$ is a symmetry variable associated to the translational broken symmetry of the smectic, its dynamic equation will be of the form
\begin{eqnarray*}
    \partial_t u + \mathbf{v}\cdot\nabla u + Y &=& 0
\end{eqnarray*}
where $Y$ is a quasi-current, which we write as $Y = Y^R + Y^D$. Since there is no restriction on $\mathbf{t}^R$ to be solenoidal, we require $\dot{u} = 0$ for reversibility, which implies that $Y^R = 0$. For deriving the irreversible quasi-current $Y^D$ that satisfies $\dot{s} > 0$, we propose a dissipation function $R$. This is a bilinear expression function, a positive definite form of the thermodynamic forces, from which irreversible currents (and quasi-currents) can be derived by taking derivatives. Accounting for restrictions on symmetry \cite{brand2001macroscopic,pleiner1996hydrodynamics}, using Einstein notation we find the following bilinear form
\begin{eqnarray*}
    R &=& \frac{1}{2}\Gamma\partial_i h_i \partial_j h_j 
    + c_j \partial_i h_i \partial_j \theta 
    + \frac{1}{2}\kappa_{ij}\partial_i\theta\partial_j\theta
    +\frac{1}{2}\eta_{ijkl}\partial_i v_j \partial_k v_l \;.
\end{eqnarray*}
where $\mathbf{h}$ is the molecular field defined by $\mathbf{h} = \partial f / \partial\nabla u$. If the temperature field is kept at a constant uniform value, we find
\begin{eqnarray*}
    Y^D &=& - \frac{\partial R}{\partial \nabla\cdot\mathbf{h}} 
    \;=\; -\Gamma \nabla\cdot\mathbf{h}\;, \quad \textrm{so that}\quad \dot{u} 
    \;=\; \Gamma \nabla\cdot\left(\frac{\partial e}{\partial \nabla u}\right) \; .
\end{eqnarray*}

The viscous part of the stress tensor is derived in the same way as in Sec. \ref{sec:comp}, written in terms of the viscosity tensor as $\mathbf{T}^D = {\boldsymbol \eta} : \nabla \mathbf{v}$. From the derived expression, we obtain the following system of governing equations 
\begin{eqnarray}
    \dot{\rho} &=& 
    -\rho\, \nabla \cdot \mathbf{v}\; ,
    \label{eq:u-bm}
    \\[2mm]
    \rho\dot{\mathbf{v}} &=& -\nabla p + 
    \nabla\cdot\bigg(-\nabla u \otimes \frac{\partial e}{\partial \nabla u}\bigg) 
    + \nabla\cdot\mathbf{T}^D\; ,
    \label{eq:u-blm}
    \\[2mm]
    \dot{u} &=& \Gamma \nabla\cdot\bigg(\frac{\partial e}{\partial \nabla u}\bigg) \quad.
    \label{eq:u-psi}
\end{eqnarray}

The commonly adopted elastic energy in terms of the configuration variable \cite{kats2012fluctuational}, for weakly distorted smectic layers with normal $\hat{k}$ in the undistorted configuration, is
\begin{eqnarray}
    \mathcal{E}_u &=& \frac{1}{2}\int_\Omega
    \Big[B(\partial_z u)^2+K_1(\partial_x^2u+\partial_y^2u)^2 \Big]
    d\mathbf{x} \; .
\end{eqnarray}

While we assumed $e = e(s,\nabla u)$, with this choice of energy we have $e = e(s,\partial_z u, \nabla^2_\perp u)$. Similarly to Chaikin and Lubensky \cite{chaikin2000principles}, we account for this difference and compute the molecular field as a functional derivative of $\mathcal{E}_u$ with respect to $\nabla u$, so that we find
\begin{eqnarray}
    \mathbf{h} &=& \frac{\partial e}{\partial \nabla u} 
    \;=\; B\,\partial_z u\,\hat{k} - K_1 \nabla_\perp\nabla^2_\perp u \;.
\end{eqnarray}

Therefore, we obtain the following equation for the configuration variable
\begin{eqnarray}
    \dot{u} &=& \Gamma\, B \partial^2_z u - \Gamma K_1 \nabla^4_\perp u \;,
\end{eqnarray}
where the first and second terms on the right hand side are associated to the permeation and undulation modes, respectively. In de Gennes and Prost notation, $\Gamma = \lambda_p$ is the permeation constant (see Eq. (8.37) in Ref. \cite{degennes1995physics}). Therefore, the balance of linear momentum, dynamic equation for $u$ and balance of mass obtained in this section agree with established results from the literature \cite{degennes1995physics,chaikin2000principles}.

It is also easily shown that our derived reversible stress is the same one found in the previous references. For instance, if we have $\nabla u \sim \hat{k}$, then
\begin{eqnarray*}
    \mathbf{T}^R &=& -\nabla u \otimes \frac{\partial e}{\partial \nabla u} \; ,
    \quad \textrm{so that} \quad T_{zx} \;=\; K_1 \partial_x\nabla^2_\perp u \;,
\end{eqnarray*}
which is the same as in Eq. (8.7) from de Gennes and Prost \cite{degennes1995physics}.


\bibliography{interface}

\end{document}